\documentclass[aps,prd,nofootinbib,showpacs]{revtex4}

\usepackage{graphicx}

\begin{document}
%
%
\def\ov{\over}
\def\beqa{\begin{eqnarray}}
\def\eeqa{\end{eqnarray}}
\def\beq{\begin{equation}}
\def\eeq{\end{equation}}
\def\p{\partial}
\def\vp{\varphi}
\def\n{\nabla}
\def\b{\beta}
\def\d{\delta}
\def\O{\Omega}
\def\ulap{\underline\Delta}
\def\unab{\overline\nabla}
\def\spose#1{\hbox to 0pt{#1\hss}}
\def\lta{\mathrel{\spose{\lower 3pt\hbox{$\mathchar"218$}}
     \raise 2.0pt\hbox{$\mathchar"13C$}}}
\def\gta{\mathrel{\spose{\lower 3pt\hbox{$\mathchar"218$}}
     \raise 2.0pt\hbox{$\mathchar"13E$}}}
\def\wg#1{\mbox{\boldmath ${#1}$}}
\def\w#1{\bf #1}

\title{Quasiequilibrium sequences of synchronized and
	irrotational binary neutron stars in general relativity.
	III. Identical and different mass stars with $\gamma=2$}

\author{Keisuke Taniguchi}

\affiliation{Department of Earth Science and Astronomy,
	Graduate School of Arts and Sciences, \\
	University of Tokyo, Komaba, Meguro, Tokyo 153-8902, Japan
}

\author{Eric Gourgoulhon}

\affiliation{Laboratoire de l'Univers et de ses Th\'eories,
	FRE 2462 du CNRS, Observatoire de Paris,
	F-92195 Meudon Cedex, France
}

\date{19 September 2002}

\begin{abstract}

We present the first computations of quasiequilibrium binary neutron stars
with different mass components in general relativity,
within the Isenberg-Wilson-Mathews approximation.
We consider both cases of synchronized rotation and irrotational motion.
A polytropic equation of state is used with the adiabatic index $\gamma=2$.
The computations have been performed for the following combinations of
stars:
$(M/R)_{\infty,\rm ~star~1}$ vs.
$(M/R)_{\infty,\rm ~star~2} = 0.12 ~{\rm vs.}~ (0.12, ~0.13, ~0.14),
~0.14 ~{\rm vs.}~ (0.14, ~0.15, ~0.16),
~0.16 ~{\rm vs.}~ (0.16, ~0.17, ~0.18),
~{\rm and} ~0.18 ~{\rm vs.}~ 0.18$,
where $(M/R)_{\infty}$ denotes the compactness parameter
of infinitely separated stars of the same baryon number.
It is found that for identical mass binary systems
there is no turning point of the binding energy (ADM mass)
before the end point of the sequence (mass shedding point)
in the irrotational case,
while there is one before the end point of the sequence
(contact point) in the synchronized case.
On the other hand, in the different mass case,
the sequence ends by the tidal disruption of the less massive star
(mass shedding point).
It is then more difficult to find a turning point in the ADM mass.
Furthermore, we find that the deformation of each star depends
mainly on the orbital separation and the mass ratio
and very weakly on its compactness.
On the other side, the decrease of the central energy density depends
on the compactness of the star and not on that of the companion.

\end{abstract}



\maketitle

\section{Introduction} \label{s:intro}

Coalescing binary neutron stars are expected to be
one of the most promising sources of gravitational waves
that could be detected by the ground based, kilometer size
laser interferometers such as the Laser Interferometric Gravitational
wave Observatory (LIGO), VIRGO, GEO600,
and TAMA300 \footnote{For TAMA300, data taking has started in
summer 1999, and the first results of the data analysis
have been published \cite{Tagoshi01}.},
and also are considered as one of candidates of
gamma-ray burst source \cite{NaraPP92}.

Due to the emission of gravitational radiation,
binary neutron stars decrease their orbital separations and finally merge.
When discussing the evolution of the system,
it is convenient to separate it into three phases.
The first one is the {\em inspiraling phase}
in which the orbital separation is much larger
than the neutron star radius,
and the post-Newtonian (PN) expansion constitutes an excellent approximation.
Recently, two groups succeeded in deriving the 3PN equation of motion
of point-mass binary systems \cite{DamourJS,BlanchetG},
and the equivalence of the results between these groups is shown in
\cite{DamourJS01,AndradeBF01}.
The second stage is the {\em intermediate phase} in which
the orbital separation becomes only a few times larger than
the radius of a neutron star,
so that hydrodynamics as well as general relativity play an important role.
In this phase, since the shrinking time of the orbital radius
due to the emission of gravitational waves is still larger
than the orbital period,
it is possible to approximate the state as quasiequilibrium.
The final stage is the {\em merging phase}
in which the two stars coalesce dynamically.
As in the intermediate phase,
since hydrodynamics as well as general relativity play an important role,
fully relativistic hydrodynamical treatments are required in this phase
which therefore pertains to the field of numerical relativity.
The first successful computations of the evolution of binary neutron stars
from their innermost stable circular orbits (ISCO) to black hole or
massive neutron star formation
have been performed by Shibata and Uryu \cite{Shiba99,ShibaU00}.
Other efforts are presented in \cite{OoharN99,FontGIMRSSST02}.

The present article belongs to the series of works \cite{BonazGM99a},
\cite{GourGTMB01} (hereafter Paper I), \cite{TanigGB01} (hereafter Paper II),
\cite{TanigG02}, devoted to the intermediate phase.
This stage is interesting because we may get informations about
equation of state of neutron stars through the ISCO \cite{FaberGRT02}.
Furthermore, it is important from a numerical point of view
since it provides initial data for the dynamical simulation
in the merging phase \cite{Shiba99,ShibaU00}.
Then numerous theoretical efforts are devoted in this phase
including (semi-)analytic Newtonian \cite{LaiRS93,TanigN00}
and post-Newtonian \cite{LombaRS97,TanigS97,ShibaT97,Tanig99} approaches
and numerical Newtonian \cite{TanigGB01,TanigG02,HachiE84a,HachiE84b,UryuE98},
post-Newtonian \cite{Shiba96}, and general relativistic
\cite{BonazGM99a,GourGTMB01,BaumgCSST97,MarroMW98,MarroMW99,UryuE00,UryuSE00,UsuiUE99,UsuiE02}
ones.
Among these studies, we concentrate on numerical computations
in the general relativistic framework.
Kochaneck \cite{Kocha92} and Bildsten and Cutler \cite{BildsC92}
have shown that the gravitational-radiation driven evolution is
too rapid for the viscous forces to synchronize the spin of each
neutron star with the orbit as they do for ordinary stellar binaries.
Rather, the viscosity is negligible and
the fluid velocity circulation (with respect to some inertial frame)
is conserved in these systems.
Provided that the initial spins are not in the millisecond regime,
this means that close binary configurations are well approximated
by zero vorticity (i.e. {\em irrotational}) states.
Formulations for irrotational binaries in general relativity
have been developed by several authors
\cite{BonazGM97,Asada98,Shiba98,Teuko98}
(see Appendix~A of Paper~I for a comparison between them).
In addition to the quasiequilibrium and irrotational assumptions,
we use the approximation of a conformally flat spatial 3-metric,
introduced by Isenberg \cite{IsenbN80}
and Wilson and Mathews \cite{WilsoM89} (hereafter IWM approximation;
see Sec. IV.C of \cite{FriedUS02} and
Sec. III.A of Paper~I for a discussion).

Until now, several groups have produced quasiequilibrium
configurations of binary neutron stars, as listed above.
Among them, there are results for synchronized \cite{BaumgCSST97,MarroMW98}
and irrotational \cite{BonazGM99a,GourGTMB01,MarroMW99,UryuE00,UryuSE00}
rotation states in the general relativistic framework,
within the IWM approximation,
and those for synchronized one within some axisymmetric approximation
\cite{UsuiUE99,UsuiE02}.
However they all deal with {\em identical} star binaries and
calculations for {\em different} mass binary system
in general relativity have not been performed yet
(in Newtonian theory they have been performed
in the synchronized case \cite{HachiE84b,TanigG02},
and recently in the irrotational one \cite{TanigG02}).

We will present here the first numerical results of
relativistic binary systems composed of different mass stars.
In addition, we give results for systems of identical stars
which extend those already presented in Paper~I.
As a first step in the study of different mass systems,
we restrict our computations to a polytropic equation of state,
with the adiabatic index $\gamma=2$.

The plan of the article is as follows.
A brief overview of our method is given in Sec.~\ref{s:method}.
In Sec.~\ref{s:tests}, new tests of the numerical code are shown,
which were not given in Papers~I and II.
We present the numerical results in Sec.~\ref{s:results},
and discuss them in Sec.~\ref{s:discussion}.
Finally Sec.~\ref{s:summary} is devoted to the summary.

Throughout this article, we adopt geometrical units: $G=c=1$,
where $G$ and $c$ denote the gravitational constant and speed of light,
respectively.

\section{Method} \label{s:method}

The reader is referred to Secs.~II, III, and IV
of Paper~I for the complete set of equations
governing perfect fluid binary stars under the assumption of
quasiequilibrium and conformally flat spatial metric,
as well as the numerical method which we use.
We simply give here some outline of the method.

Let us summarize the assumptions expressed in Sec.~\ref{s:intro}.
The first one is {\em quasiequilibrium},
resulting from the fact that the time scale of orbital shrinking is
larger than that of the orbital revolution until the ISCO.
This assumption is taken into account by demanding
that the spacetime is endowed with a {\em helical Killing vector}
\beq
  \wg{l} = {\p \over \p t} +\O {\p \over \p \vp},
\eeq
where $\O$ is the orbital angular velocity and the vectors
$\p / \p t$ and $\p / \p \vp$ are associated with the
time coordinate $t$ and the azimuthal coordinate $\varphi$
of an asymptotical inertial observer.
The second assumption regards the matter stress-energy tensor,
which we assume to have the {\em perfect fluid} form:
\beq
  T_{\mu \nu} = (e+p) u_{\mu} u_{\nu} +p g_{\mu \nu},
\eeq
where $e$ denotes the fluid proper energy density,
$p$ the fluid pressure, $u_{\mu}$ the fluid 4-velocity,
and $g_{\mu \nu}$ the spacetime metric.
The third assumption concerns the rotation state of the binary system.
Although the realistic rotation state will be an irrotational one
(cf. Sec.~\ref{s:intro}),
we consider both cases of {\em synchronized} and {\em irrotational} states
in order to exhibit their differences.
The fourth assumption is on the equation of state.
For simplicity, we use a {\em polytropic} one,
\beq \label{e:eos}
  p=\kappa n^{\gamma},
\eeq
where $n$ is the fluid baryon number density and
$\kappa$ and $\gamma$ are some constants.
Here we assume that the two neutron stars have the same equation of state,
i.e., the constants $\kappa$ and $\gamma$ are identical
for both stars.
The fifth and last assumption is the IWM one, namely
a {\em conformally flat spatial metric}.
Then the full spacetime metric takes the form:
\beq
  ds^2 =-(N^2 -B_i B^i) dt^2 -2B_i dt dx^i +A^2 f_{ij} dx^i dx^j,
\eeq
$N$ being the lapse function, $B^i$ the shift vector,
$A$ the conformal factor, and $f_{ij}$ the flat spatial metric.
Under these assumptions,
the equations for fluid motion and gravitational field are presented
in Paper~I.

\begin{figure}[htb]
\vspace{0.5cm}
\begin{center}
  \includegraphics[width=8cm]{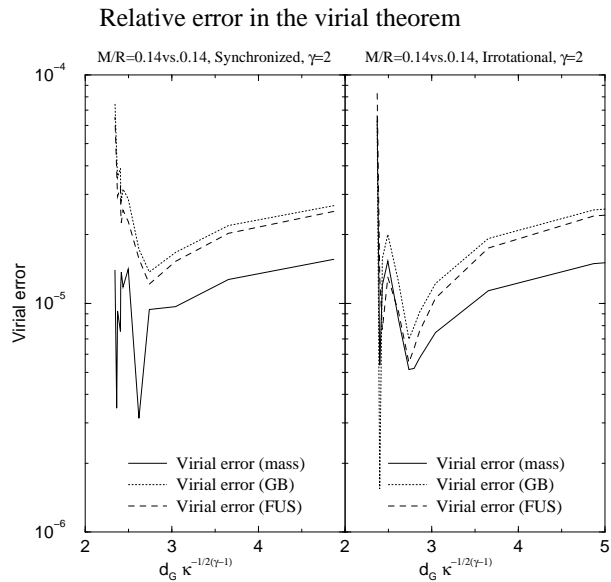}
\end{center}
\caption[]{\label{fig:virial}
Relative error in the virial theorem for
numerical solutions of relativistic quasiequilibrium binary system,
as a function of the orbital separation.
Solid, dashed, and dotted lines denote the relative errors
defined by Eq.~(\ref{eq:vemass}), (\ref{eq:vefus}) and
(\ref{eq:vegb}), respectively.
Left panel is for synchronized binaries, and
the right one for irrotational ones.
Both of them are drawn for identical mass binary systems
with the compactness $M/R=0.14$ vs. 0.14.
}
\end{figure}%

\section{Code tests} \label{s:tests}

In this series of studies \cite{BonazGM99a,GourGTMB01,TanigGB01,TanigG02},
we use a multidomain spectral method with surface-fitted
coordinates \cite{BonazGM98,BonazGM99b,GrandBGM01}.
This method is implemented by a numerical code constructed upon
the C++ library {\sc Lorene} \cite{lorene}.
Numerous tests of the code have been already presented in Papers~I
and II. In particular, the comparison with results
from other groups \cite{BaumgCSST97,UryuE00} has been given in
Sec.~V.D of Paper~I for the identical mass case.
We will give here new tests based on the relative error
in a general relativistic generalization of the virial theorem.

In Newtonian gravity, the virial theorem has proved to be useful to check the
global error in numerical solutions for stationary fluid systems.
A general relativistic version of the virial theorem has been
obtained by Gourgoulhon and Bonazzola \cite{GourgB94} for
stationary spacetimes. It has been recently extended to
binary star spacetimes within the IWM approximation by
Friedman, Uryu and Shibata \cite{FriedUS02}
\footnote{The spacetime generated by a binary system is
not stationary, due to gravitational radiation.
However, within the IWM approximation, the gravitational
radiation is neglected in the global spacetime dynamics,
so that one is able to recover the virial theorem.}.
The virial relation is equivalent to
\beq
  M_{\rm ADM} - M_{\rm Komar} =0,
\eeq
where $M_{\rm ADM}$ is the ADM mass:
\beq  \label{e:M_ADM}
  M_{\rm ADM} = -{1 \over 2\pi} \oint_{\infty} \unab^i A^{1/2} dS_i,
\eeq
and $M_{\rm Komar}$ is a Komar-type mass, defined by
\beq
  M_{\rm Komar} = {1 \over 4\pi} \oint_{\infty} \unab^i N dS_i.
\eeq
The virial theorem obtained by Friedman, Uryu and Shibata \cite{FriedUS02}
(see also Eq. (5.7) of Ref. \cite{ShibaU01}) writes
\beqa
  VE(FUS) &=&\int \Bigl[ 2N A^3 S +{3 \over 8\pi} N A^3 K_i^j K_j^i
	+{1 \over 4\pi} N A (\unab_i \b \unab^i \b -\unab_i \nu \unab^i \nu)
	\Bigr] \nonumber \\
	&=&0. \label{eq:ve_fus}
\eeqa
By a straightforward manipulation, this integral can be recast in
the form of the virial theorem as obtained by Gourgoulhon and Bonazzola
\cite{GourgB94}:
\beqa
  VE(GB) &=&\int \Bigl[ 2A^3 S +{3 \over 8\pi} A^3 K_i^j K_j^i
	+{1 \over 4\pi} A (\unab_i \b \unab^i \b -\unab_i \nu \unab^i \nu
	-2 \unab_i \b \unab^i \nu) \Bigr] \nonumber \\
	&=&0 . \label{eq:ve_gb}
\eeqa
Let us stress that the above identity has been derived by
Gourgoulhon and Bonazzola \cite{GourgB94} only for stationary
spacetimes and that its validity for IWM spacetimes with
helical Killing vector has been obtained by Friedman,
Uryu and Shibata \cite{FriedUS02}.

As error indicators of our numerical solutions,
we have evaluated the quantities
\beq
  \Bigl| {M_{\rm ADM} - M_{\rm Komar} \over M_{\rm ADM}} \Bigr|,
	\label{eq:vemass}
\eeq
\beq
  \Bigl| {VE(FUS) \over M_{\rm ADM}} \Bigr|, \label{eq:vefus}
\eeq
\beq
  \Bigl| {VE(GB) \over M_{\rm ADM}} \Bigr|. \label{eq:vegb}
\eeq
The results are presented in Fig. \ref{fig:virial}
as a function of the orbital separation.
One can see from this figure that the typical virial error is
$\sim 1 \times 10^{-5}$ through the sequence.

\begin{figure}[htb]
\vspace{0.5cm}
\begin{center}
  \includegraphics[width=8cm]{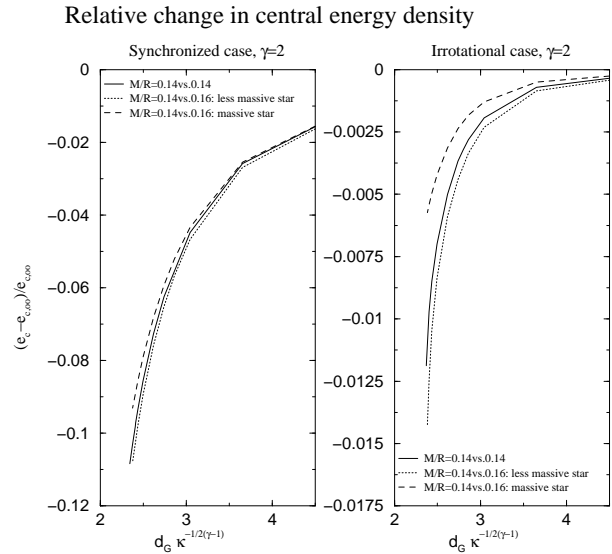}
\end{center}
\caption[]{\label{fig:change}
Relative change in central energy density
as a function of the orbital separation between the centers of mass
of each star.
Left (resp. right) panel is for synchronized (resp. irrotational)
binaries with an adiabatic index $\gamma=2$.
Solid line denotes the identical mass case $M/R=0.14$ vs. $0.14$.
The results of the different mass case $M/R=0.14$ vs. $0.16$
are shown as dotted (less massive star) and dashed (massive star) lines.
}
\end{figure}%

\section{Numerical results} \label{s:results}

In this section, we present the numerical results
for quasiequilibrium sequences of constant baryon number
binaries (evolutionary sequences),
for both synchronized and irrotational cases and for the
adiabatic index $\gamma=2$.
Although the interesting case is the irrotational one,
as mentioned in Sec. \ref{s:intro},
we have computed configurations for both rotation states
in order to compare their properties.
The number of computational domains is 4 or 5 for each star
and the space around it.
The domains have an onion-like structure, being
centered on each star:
the innermost domain covers the star,
two or three domains (shells) are placed outside it,
and the external domain extends to infinity thanks to some compactification
(cf. Sec.~IV.A of Paper~I for all details).
The total number of domains is therefore $2\times4 = 8$ or
$2\times 5=10$. We use 8 domains for close orbital separation and
10 domains for large orbital separation.
The numbers of collocation points in each domains are
$N_r \times N_{\theta} \times N_{\varphi} =33 \times 25 \times 24$ or
$25 \times 17 \times 16$,
where $N_r$, $N_{\theta}$, and $N_{\varphi}$ denote the number of
collocation points of the radial, the polar, and the azimuthal directions,
respectively.

We parametrize the different mass systems by the values of
$(M/R)_{\infty,\rm ~star~1}$ vs. $(M/R)_{\infty,\rm ~star~2}$,
where $(M/R)_\infty$ denotes the compactness of the isolated
spherical star with the same baryon number.

First of all, the relative changes in central energy density
for the cases of $M/R=0.14$ vs. 0.14 and
0.14 vs. 0.16 are depicted in Fig. \ref{fig:change}
as a function of the orbital separation.
(Hereafter we abbreviate $(M/R)_{\infty,\rm ~star~1}$ vs.
$(M/R)_{\infty,\rm ~star~2}$ as $M/R$ for simplicity.)
One can see from these figures that the central energy density
decreases monotonically in both synchronized and irrotational cases
\footnote{In Paper I (Table~III), as well as
in \cite{BonazGM99a} (Fig.~1), there was a slight increase in the
central energy density for irrotational binaries
at the large and medium separations.
However, from Paper~II
we have improved the method to determine the surface of the star,
which leads to a better accuracy in the present work.}.
The decrease of the central energy density for the synchronized case
is about 10~\% or more at the end of the sequence,
while for the irrotational case, it is only about 1~\%.

\begin{figure}[htb]
\vspace{0.5cm}
\begin{center}
  \includegraphics[width=7cm]{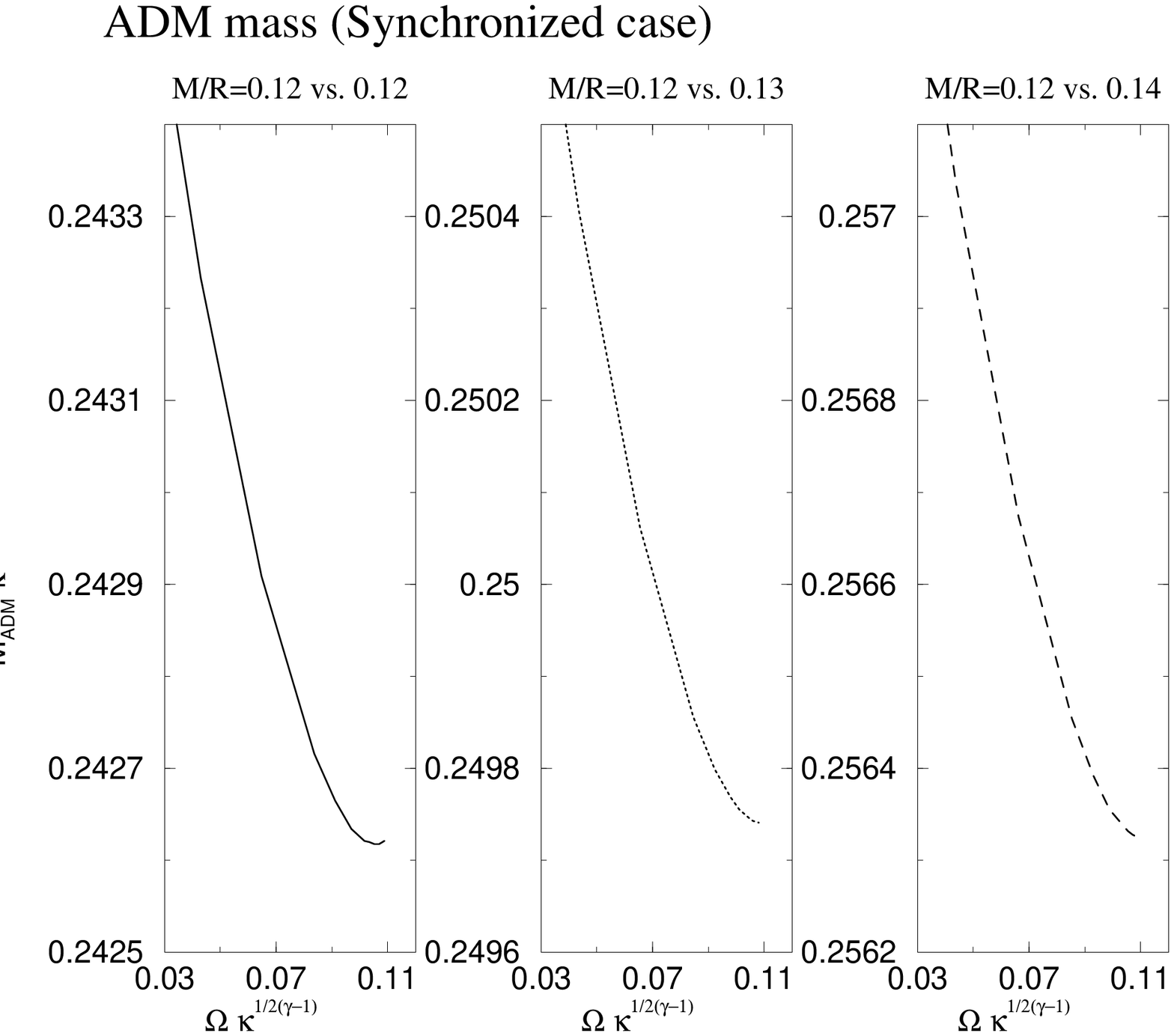}
  \hspace{20pt} \includegraphics[width=7cm]{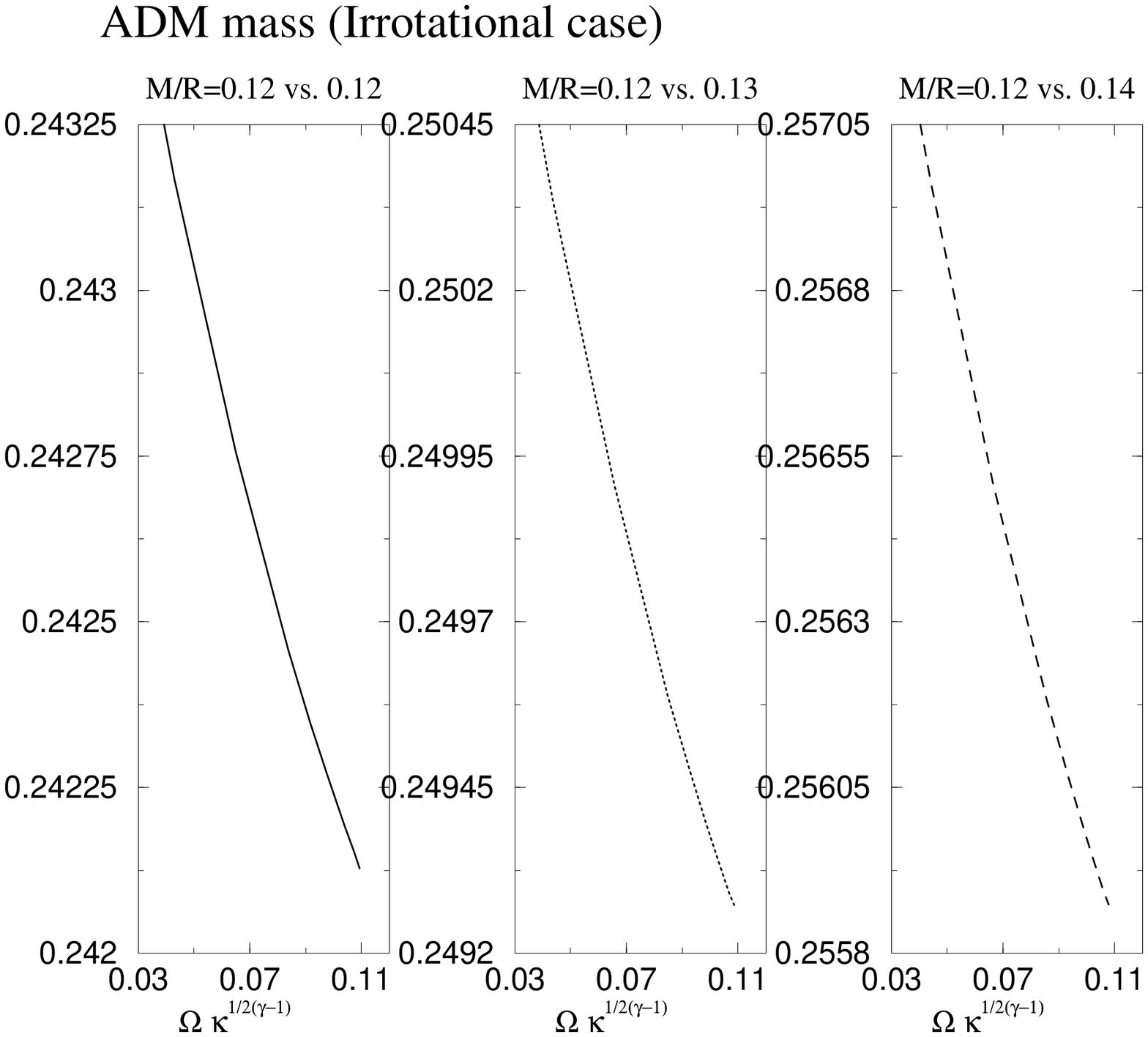}
\end{center}
\caption[]{\label{fig:adm_mass_12}
ADM mass of the binary system
as a function of the orbital angular velocity.
Left (resp. right) panel is for synchronized
(resp. irrotational) binaries.
In each panel, the three sub-panels correspond to
the compactness of $M/R=0.12$ vs. 0.12, 0.12 vs. 0.13,
and 0.12 vs. 0.14 from the left to the right.
}
\end{figure}%

\begin{figure}[htb]
\vspace{0.5cm}
\begin{center}
  \includegraphics[width=7cm]{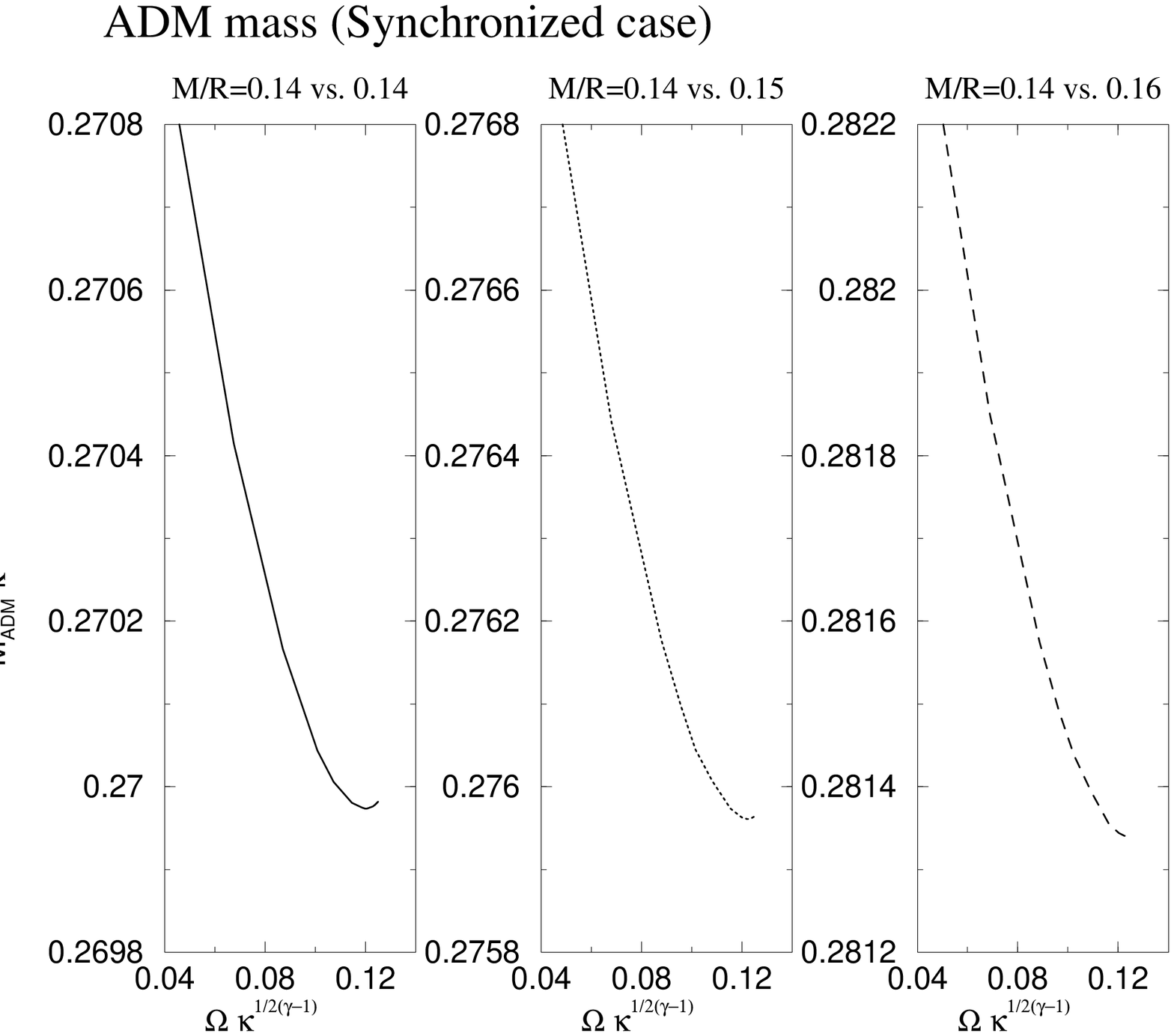}
  \hspace{20pt} \includegraphics[width=7cm]{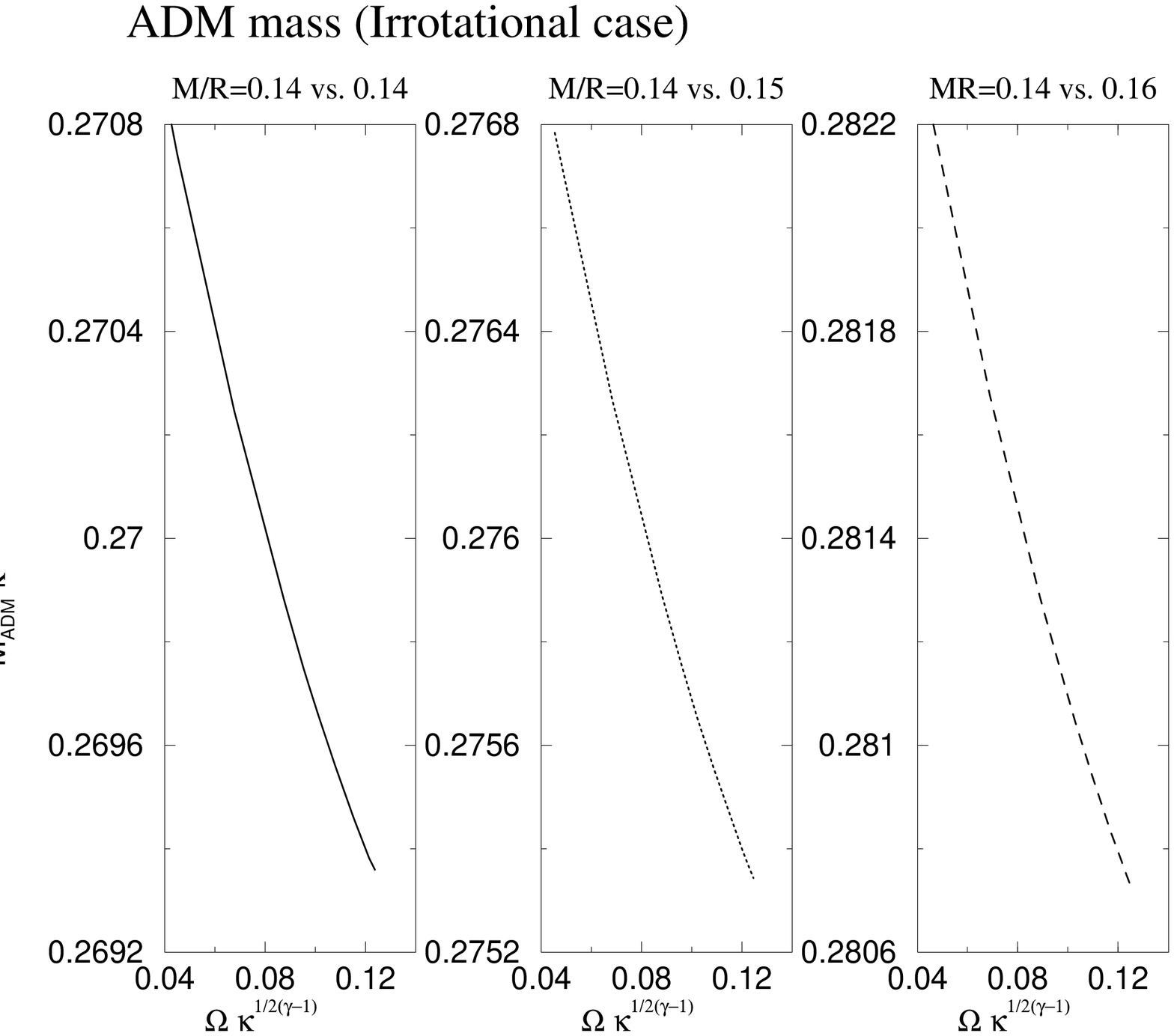}
\end{center}
\caption[]{\label{fig:adm_mass_14}
Same as Fig.~\ref{fig:adm_mass_12}, but
for the compactness of $M/R=0.14$ vs. 0.14, 0.14 vs. 0.15,
and 0.14 vs. 0.16 from the left to the right in each panel.
}
\end{figure}%

\begin{figure}[htb]
\vspace{0.5cm}
\begin{center}
  \includegraphics[width=7cm]{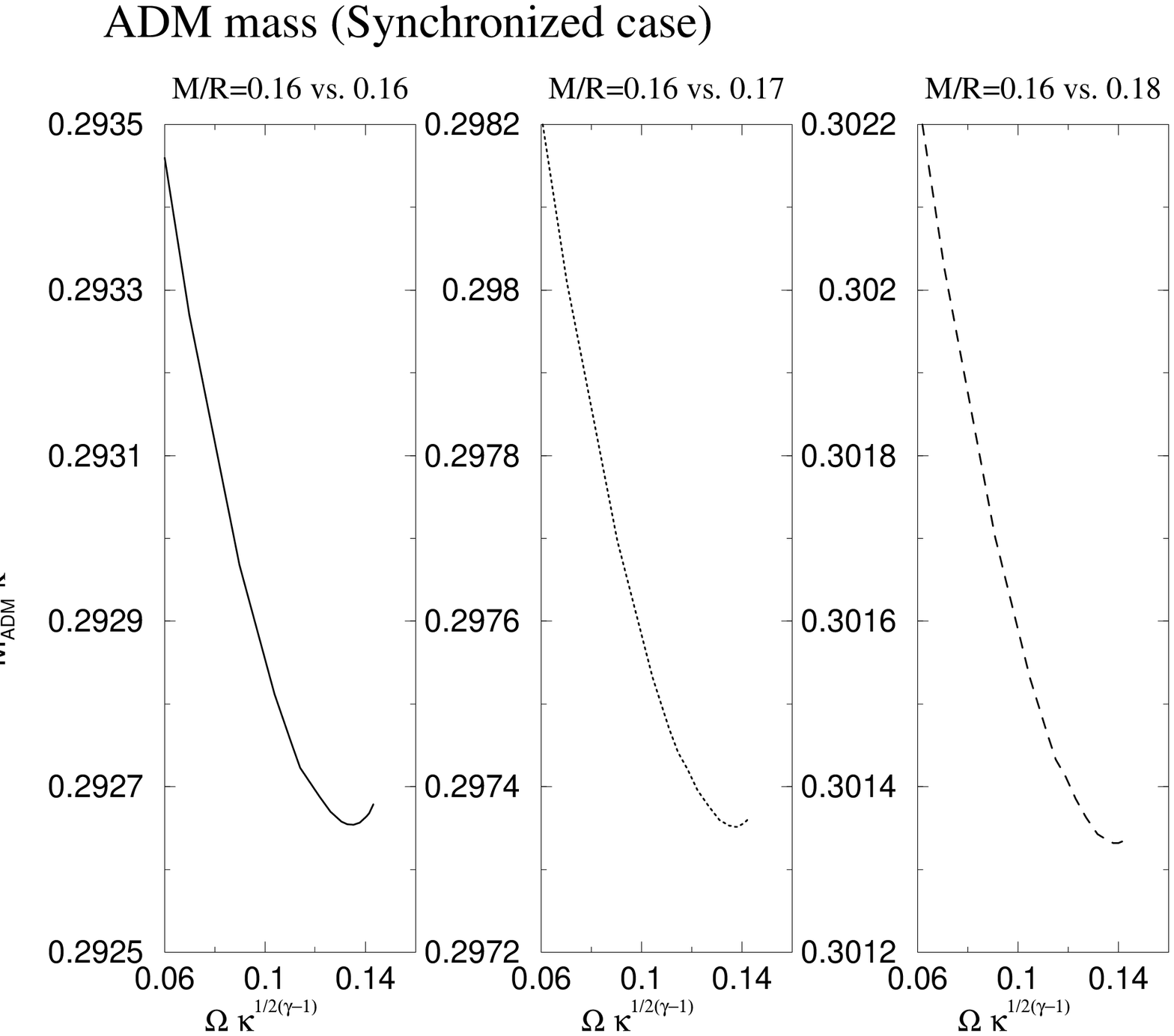}
  \hspace{20pt} \includegraphics[width=7cm]{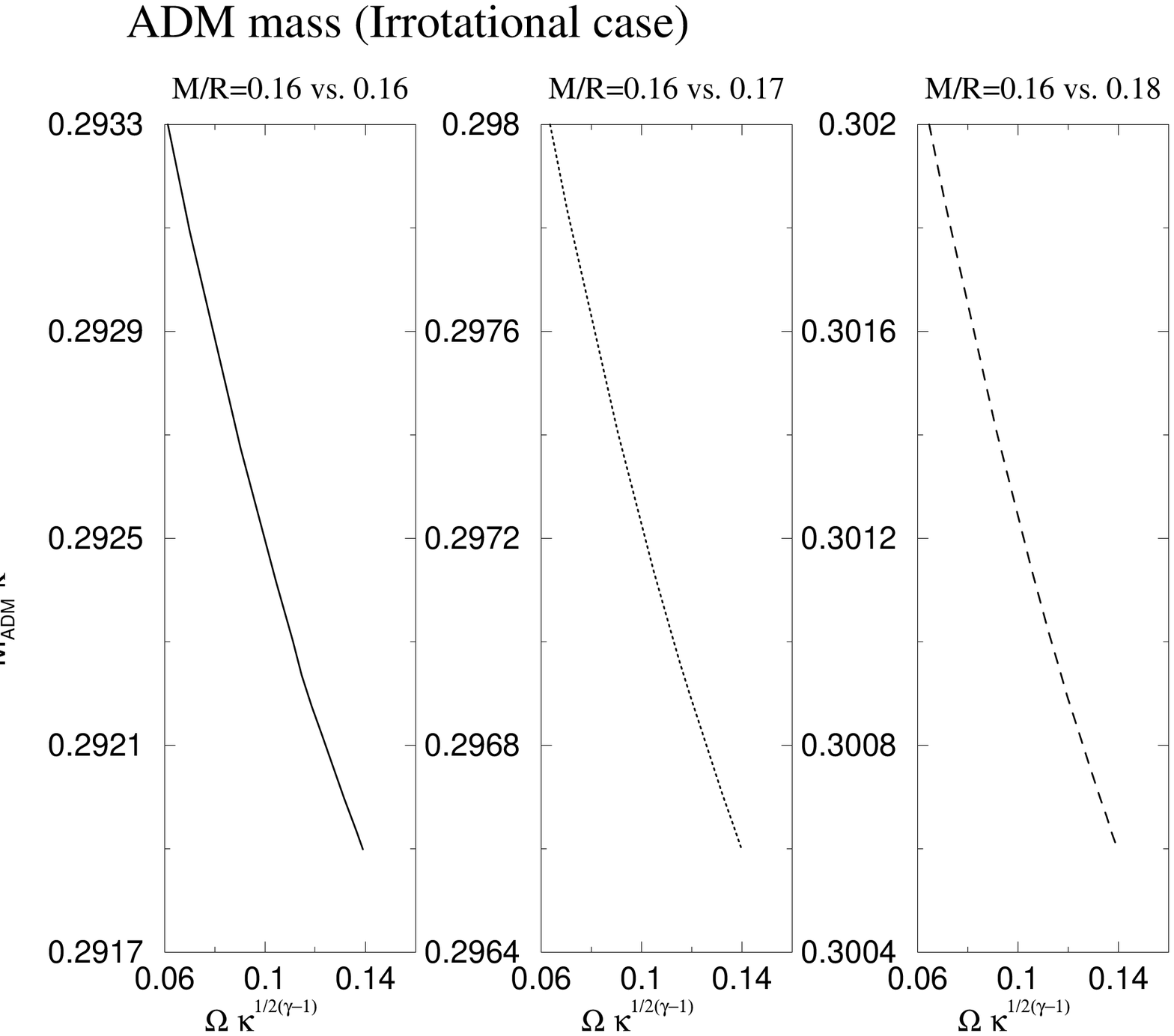}
\end{center}
\caption[]{\label{fig:adm_mass_16}
Same as Fig.~\ref{fig:adm_mass_12}, but
for the compactness of $M/R=0.16$ vs. 0.16, 0.16 vs. 0.17,
and 0.16 vs. 0.18 from the left to the right in each panel.
}
\end{figure}%

Next, the ADM mass of the binary system [Eq.~(\ref{e:M_ADM})]
is shown in Figs. \ref{fig:adm_mass_12} -- \ref{fig:adm_mass_16}
as a function of the orbital angular velocity, for
different mass ratios and different states of rotation.
One can clearly see the turning point in the synchronized cases
for identical mass binaries.
For more compact stars the turning points appear relatively earlier
than those of less compact ones.
On the other hand, we do not find any turning point
in irrotational binaries.
Moreover, it becomes more difficult for the turning point to appear
even in the synchronized case
when the mass difference in the binary system increases.

\begin{figure}[htb]
\vspace{0.5cm}
\begin{center}
  \includegraphics[width=8cm]{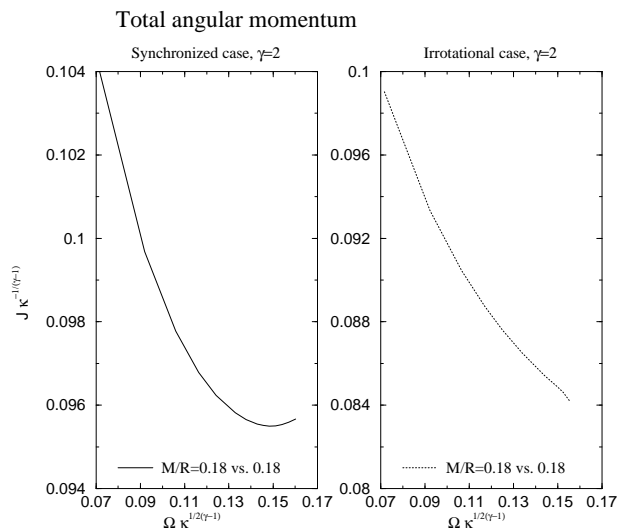}
\end{center}
\caption[]{\label{fig:ang_mom}
Total angular momentum of the binary system
as a function of the orbital angular velocity
for the compactness $M/R=0.18$ vs. 0.18.
Left panel is for the synchronized case and right for the irrotational one.
}
\end{figure}

In Fig. \ref{fig:ang_mom}, we show the total angular momentum
of the binary system as a function of the orbital angular velocity
for the compactness $M/R=0.18$ vs. 0.18.
For synchronized binaries, the turning point of the total angular
momentum coincides with that of the ADM mass within the numerical accuracy.

\begin{figure}[htb]
\vspace{0.5cm}
\begin{center}
  \includegraphics[width=8cm]{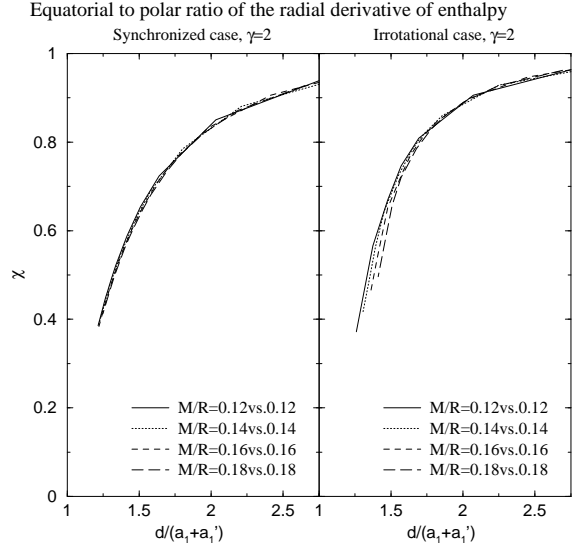}
\end{center}
\caption[]{\label{fig:chi}
Equatorial to polar ratio of the radial derivative of the enthalpy
as a function of the orbital separation.
Left (resp. right) panel is for synchronized (resp. irrotational)
binaries.
Solid, dotted, dashed, and long-dashed lines denote the cases of
the compactness $M/R=0.12$ vs. 0.12, 0.14 vs. 0.14, 0.16 vs. 0.16,
and 0.18 vs. 0.18, respectively.
}
\end{figure}%

\begin{figure}[htb]
\vspace{0.5cm}
\begin{center}
  \includegraphics[width=8cm]{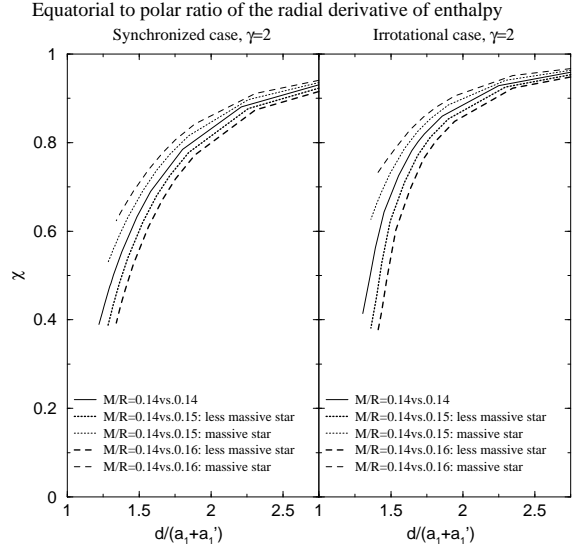}
\end{center}
\caption[]{\label{fig:chi_diff}
Same as Fig.~\ref{fig:chi}, but for different mass binaries.
Left (resp. right) panel is for synchronized (resp. irrotational)
binaries.
Solid line denotes the case of identical mass binary
with the compactness $M/R=0.14$.
Thin and thick dotted lines are for the compactness $M/R=0.14$ vs. 0.15
of the more massive and less massive stars, respectively.
Thin and thick dashed lines are for the compactness $M/R=0.14$ vs. 0.16
of the more massive and less massive stars, respectively.
}
\end{figure}%

The equatorial to polar ratios of the radial derivative
of the enthalpy are shown as a function of the orbital separation
normalized by the radius of a star to the companion star
in Figs. \ref{fig:chi} and \ref{fig:chi_diff}.
As explained in Papers~I and II,
this quantity
\beq \label{e:chi}
  \chi := {(\p H/\p r)_{\rm eq, comp} \over
	(\p H/\p r)_{\rm pole}}
\eeq
is a good dimensionless indicator to detect
the appearance of cusp at the stellar surface, which
corresponds to the mass shedding limit.
It is worth to plot the variation of $\chi$ with respect
to the orbital separation
normalized by the sum of the radius of each star toward its companion:
$d/(a_1+a_1')$,
where $d$ denotes the orbital separation between the two coordinate
center (maximum of density),
$a_1$ the radius of the less massive star in the direction
of its companion,
and $a_1'$ that of the more massive star.
Note here that the separation $d$ and the radii $a_1$ and $a_1'$ are
coordinate lengths.
If the quantity $\chi$ becomes zero, we can conclude that
the sequence ends by mass shedding.
On the other hand, if the quantity $d/(a_1+a_1')$ becomes unity,
the two stars will contact each other.
It is found from Fig.~\ref{fig:chi} that for synchronized
identical mass binaries
the sequences seem to terminate by the contact between the two stars
(the quantity $\chi$ will become zero at the same time).
The important feature of the sequences is that
they take almost the same track regardless of the compactness.
On the contrary, one can see from Fig.~\ref{fig:chi}
that for irrotational binaries the sequences seem to
terminate by a cusp point (mass shedding).
Such a behavior is enhanced for more compact cases.
In Fig.~\ref{fig:chi_diff}, we compare the quantity $\chi$
of identical mass binary of the compactness $M/R=0.14$
with that of different mass binaries
of the compactness $M/R=0.14$ vs. 0.15 and 0.14 vs. 0.16.
It is found that the sequences seem to terminate by a cusp point
of the less massive star for both synchronized and irrotational binaries
when we extrapolate the lines,
while the more massive star keeps its shape near spherical one
(see the next paragraph).
When we increase the difference in mass between two stars,
this tendency is enhanced.

It is worth to mention that as soon as the two masses
differ, even slightly,
the sequence of a synchronized binary system terminates by a cusp point,
because the line for the less massive star is always located below that
for the identical mass stars which will reach $\chi=0$
at $d/(a_1+a_1')=1$.
This implies that the line for the less massive star seems to always
reach $\chi=0$ at $d/(a_1+a_1') > 1$.
Note here that the sequences for identical mass binaries and
less massive stars in different mass binaries should reach the point
$\chi=0$. However, in our numerical method, we cannot treat a cuspy figure,
because we adapt the innermost numerical domain to the surface
of the star and the multidomain spectral method that we use
assume that the domain boundaries are smooth surfaces.
For very close configurations, it becomes therefore difficult
to make the iterative procedure converge to a sufficient level, i.e. to
have the relative difference $\delta H$ in the enthalpy field between two steps
lower than $10^{-5}$.
This explains why we stop the sequences at $\chi \simeq 0.4$.
We use $\delta H=10^{-10}$ for very large separations,
$\delta H=10^{-8} - 10^{-7}$ for medium ones,
and $\delta H=10^{-6} - 10^{-5}$ for very close ones.

\begin{figure}[htb]
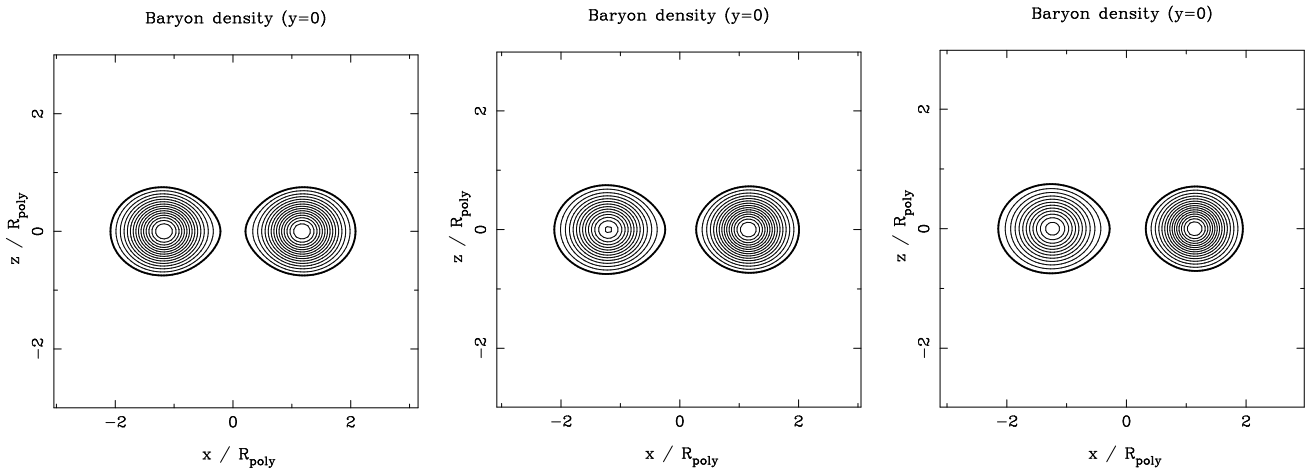

\vspace{0.5cm}
\begin{center}
  \includegraphics[width=5.5cm]{fig9a.eps}
  \hspace{5pt} \includegraphics[width=5.5cm]{fig9b.eps}
  \hspace{5pt} \includegraphics[width=5.5cm]{fig9c.eps}
\end{center}
\caption[]{\label{fig:baryon_co}
Isocontours of the baryon density for synchronized binary systems.
Left panel is for the compactness $M/R=0.14$ vs. 0.14,
the center panel $M/R=0.14$ vs. 0.15, and the right panel
$M/R=0.14$ vs. 0.16.
The orbital axis is located on $x=0$ in each panel.
}
\end{figure}%

\begin{figure}[htb]
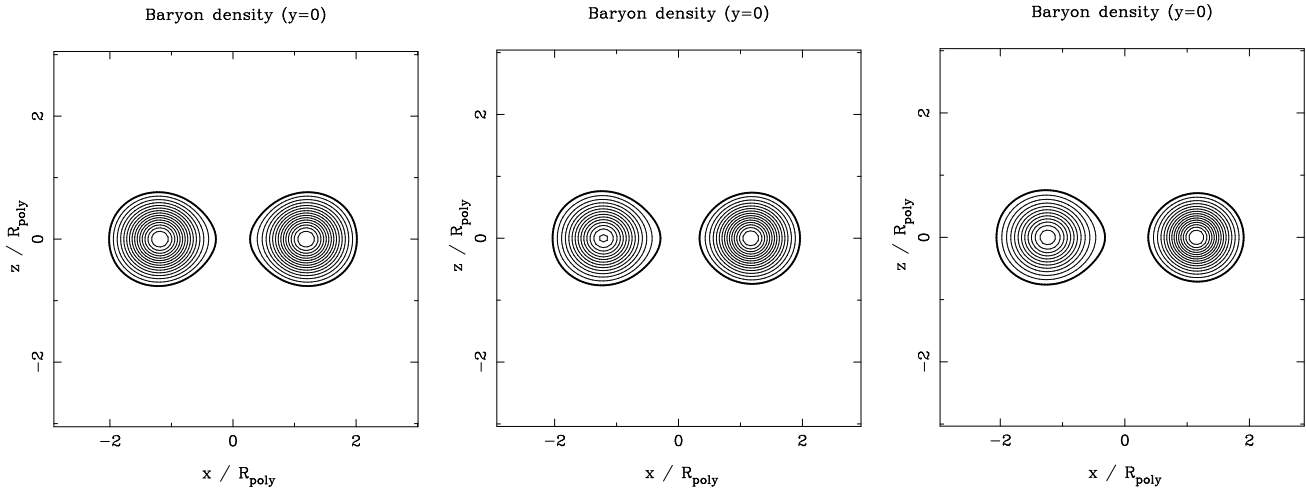

\vspace{0.5cm}
\begin{center}
  \includegraphics[width=5.5cm]{fig10a.eps}
  \hspace{5pt} \includegraphics[width=5.5cm]{fig10b.eps}
  \hspace{5pt} \includegraphics[width=5.5cm]{fig10c.eps}
\end{center}
\caption[]{\label{fig:baryon_ir}
Same as Fig.~\ref{fig:baryon_co}, but for irrotational binary systems.
}
\end{figure}%

In Figs.~\ref{fig:baryon_co} and \ref{fig:baryon_ir},
isocontours of the baryon density of various binary systems are shown.
There are three panels in each figure.
The left panel shows the result for identical mass binary
with the compactness $M/R=0.14$ vs. 0.14.
The center and right ones show the results for different mass binaries
with the compactness $M/R=0.14$ vs. 0.15, and 0.14 vs. 0.16, respectively.
In these panels, the left hand side star is the less massive star.
It is clearly seen that the less massive star is tidally deformed
and elongated while the more massive star relatively
does not deviate from the spherical shape so much
(see the right panel).
The orbital separations of these isocontours are listed at the last line
of each case in Tables below
which corresponds to the closest figure we can calculate,
except for the case of $M/R=0.14$ vs. 0.15
which we do not show in the present paper as a table.

Finally, our numerical results for constant baryon number sequences
are presented in Tables~\ref{table1} -- \ref{table4}.
We represent the orbital separation by two quantities:
\beqa
  \bar{d}_G &=& {d_G \over R_{\rm poly}}, \\
  \bar{d} &=& {d \over R_{\rm poly}} \ .
\eeqa
$d_G$ is the coordinate separation between the centers of mass of
each star (same definition as Eq. (107) of \cite{UryuE00}
or Eq. (128) of Paper I),
$d$ that between the two stellar centers defined as the maxima of
the density field, and $R_{\rm poly}$ the length
constructed from the polytropic constants $\kappa$
and $\gamma$ [Eq.~(\ref{e:eos})]
\beq
  R_{\rm poly} := \kappa^{1/2(\gamma-1)}.
\eeq
The dimensionless orbital angular velocity is defined by
\beq
  \bar{\Omega} := \Omega \kappa^{1/2(\gamma-1)},
\eeq
the dimensionless ADM mass by
\beq
  \bar{M} := M_{\rm ADM} \kappa^{-1/2(\gamma-1)},
\eeq
the dimensionless baryon mass by
\beq
  \bar{M}_{\rm B} := M_{\rm B} \kappa^{-1/2(\gamma-1)},
\eeq
and the dimensionless total angular momentum by
\beq
  \bar{J} := J \kappa^{-1/(\gamma-1)}.
\eeq
The virial errors $\bar{VE(M)}$, $\bar{VE(GB)}$ and $\bar{VE(FUS)}$
are defined by Eqs.~(\ref{eq:vemass}), (\ref{eq:vegb}) and
(\ref{eq:vefus}) respectively.
$a_1$ denotes the coordinate radius toward the companion star, $a_2$ the
radius perpendicular to it in the orbital plane and $a_3$
the radius perpendicular to the orbital plane. $\d e_{\rm c}$
is the relative change in central energy density:
$\d e_{\rm c}:=(e_{\rm c}-e_{{\rm c},\infty})/e_{{\rm c},\infty}$,
and the cusp indicator $\chi$ is defined by Eq.~(\ref{e:chi}).
In the Tables,
the superscript `` ' '' denotes the values for the more massive star.
and the symbol $\dagger$ denotes the turning point
in the ADM mass along the sequence.

\begin{figure}[htb]
\vspace{0.5cm}
\begin{center}
  \includegraphics[width=8cm]{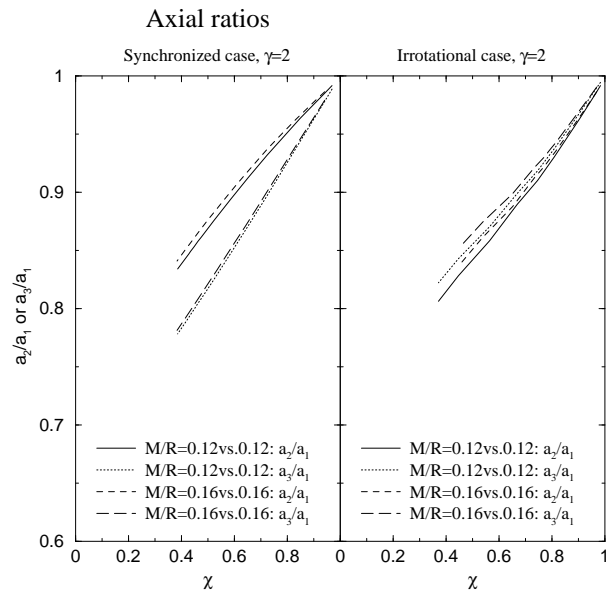}
\end{center}
\caption[]{\label{fig:axial_ratio_eq}
Axial ratios $a_2/a_1$ or $a_3/a_1$
as a function of the cusp indicator $\chi$.
Left (resp. right) panel is for synchronized (resp. irrotational)
binaries.
Solid and dotted lines denote the axial ratios $a_2/a_1$ and $a_3/a_1$
for the compactness $M/R=0.12$ vs. 0.12, respectively.
Dashed and long-dashed lines are the axial ratios $a_2/a_1$ and $a_3/a_1$
for the compactness $M/R=0.16$ vs. 0.16, respectively.
}
\end{figure}%

\begin{figure}[htb]
\vspace{0.5cm}
\begin{center}
  \includegraphics[width=8cm]{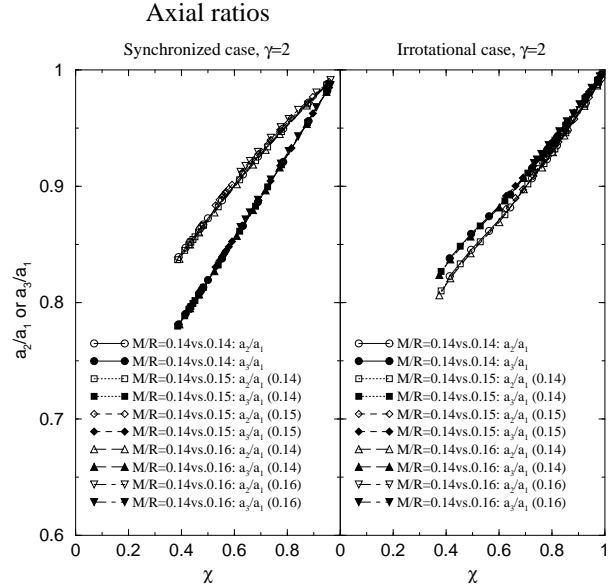}
\end{center}
\caption[]{\label{fig:axial_ratio_df}
Same as Fig.~\ref{fig:axial_ratio_eq} but for different mass binaries.
Solid line with open and filled circles denote the axial ratios
$a_2/a_1$ and $a_3/a_1$ of the compactness $M/R=0.14$ vs. 0.14, respectively.
Dotted line with open and filled squares denote the axial ratios
$a_2/a_1$ and $a_3/a_1$ of the compactness $M/R=0.14$ vs. 0.15
for less massive star, respectively.
Dashed line with open and filled diamonds denote the axial ratios
$a_2/a_1$ and $a_3/a_1$ of the compactness $M/R=0.14$ vs. 0.15
for more massive star, respectively.
Long-dashed line with open and filled triangle-up denote the axial ratios
$a_2/a_1$ and $a_3/a_1$ of the compactness $M/R=0.14$ vs. 0.16
for less massive star, respectively.
Dot-dashed line with open and filled triangle-down denote the axial ratios
$a_2/a_1$ and $a_3/a_1$ of the compactness $M/R=0.14$ vs. 0.16
for more massive star, respectively.
}
\end{figure}%

\section{Discussion} \label{s:discussion}

In this section, we discuss about the behaviors of the axial ratios,
the relative change in central energy density,
and the turning point of the synchronized sequences.

In Figs. \ref{fig:axial_ratio_eq} and \ref{fig:axial_ratio_df},
the axial ratios are shown as a function of the cusp
indicator $\chi$.
In Fig. \ref{fig:axial_ratio_eq}, we show the axial ratios $a_2/a_1$
and $a_3/a_1$ for identical mass binaries with the compactness
$M/R=0.12$ vs. 0.12 and 0.16 vs. 0.16.
One can see from this figure that each axial ratio has almost the same
track regardless of the difference of the compactness,
in particular for synchronized binary systems.
Moreover, such a coincidence of the sequence appears in the cases of
different mass binary systems (see Fig. \ref{fig:axial_ratio_df}).
In those cases, the sequences almost coincide with each other
regardless of the compactness of the companion star.
Note that the sequences of the axial ratios for more massive stars
end before $\chi=0$, while those for less massive stars should reach
the point $\chi=0$.
However we cannot treat a cuspy figure
with our numerical method, because we adapt the innermost numerical domain
to the surface of the star and the domain boundaries are assumed to be
smooth (see discussion in Sec.~\ref{s:results}).
Therefore we had to stop the sequences at around $\chi=0.4$ for the
less massive star.

\begin{figure}[htb]
\vspace{0.5cm}
\begin{center}
  \includegraphics[width=8cm]{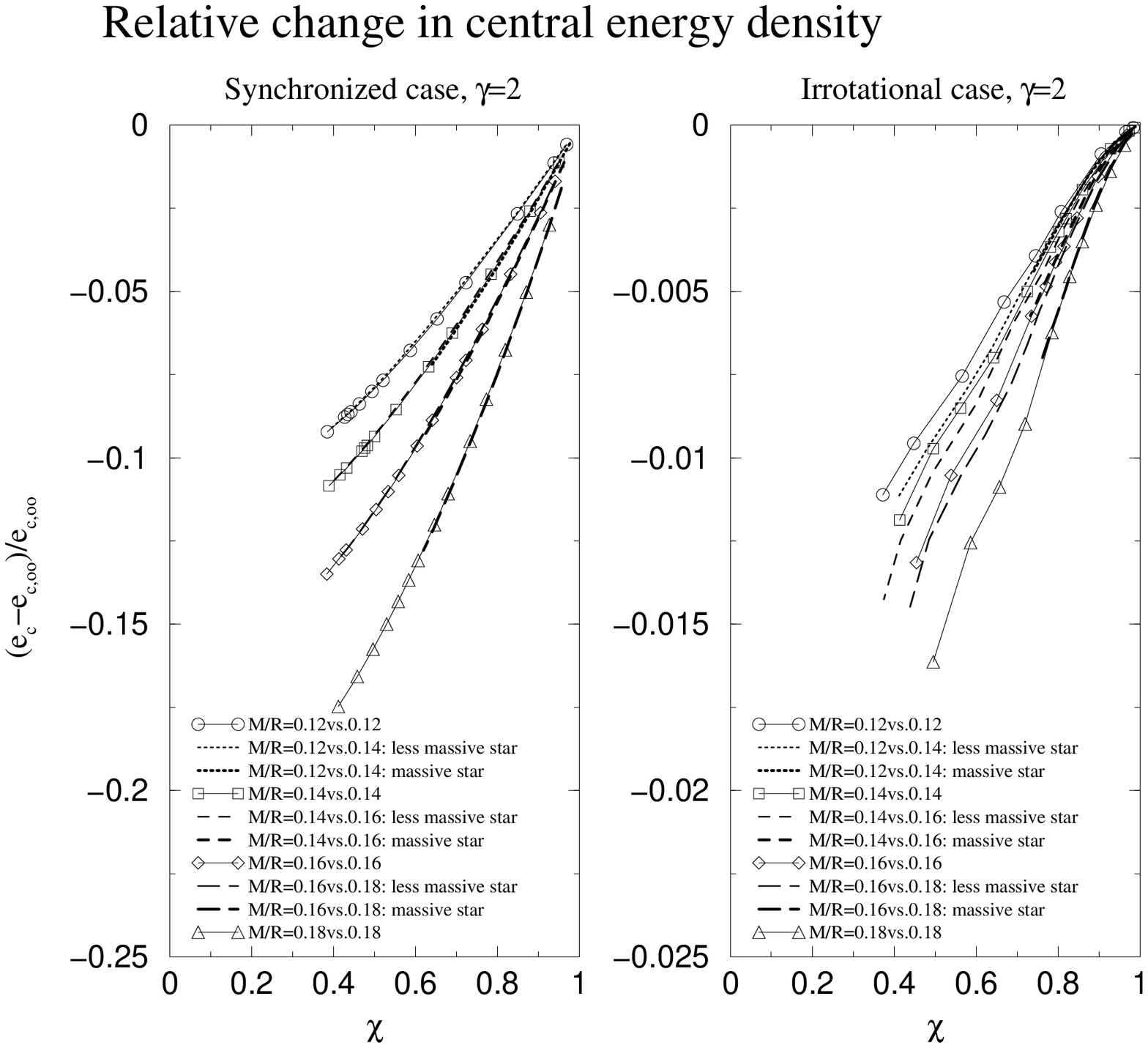}
\end{center}
\caption[]{\label{fig:dec_vs_chi}
Relative change in central energy density
as a function of the cusp indicator $\chi$.
Left (resp. right) panel is for synchronized (resp. irrotational) binaries.
Solid lines with open circle, square, diamond, and triangle denote
the cases of the identical mass binaries with the compactness
$M/R=0.12$ vs. 0.12, 0.14 vs. 0.14, 0.16 vs. 0.16, and 0.18 vs. 0.18,
respectively.
Thin and thick dotted lines are for the compactness
$M/R=0.12$ vs. 0.14 of the less massive and
more massive stars, respectively.
Thin and thick dashed lines are for the compactness
$M/R=0.14$ vs. 0.16 of the less massive and
more massive stars, respectively.
Thin and thick long-dashed lines are for the compactness
$M/R=0.16$ vs. 0.18 of the less massive and
more massive stars, respectively.
}
\end{figure}%

\begin{figure}[htb]
\vspace{0.5cm}
\begin{center}
  \includegraphics[width=8cm]{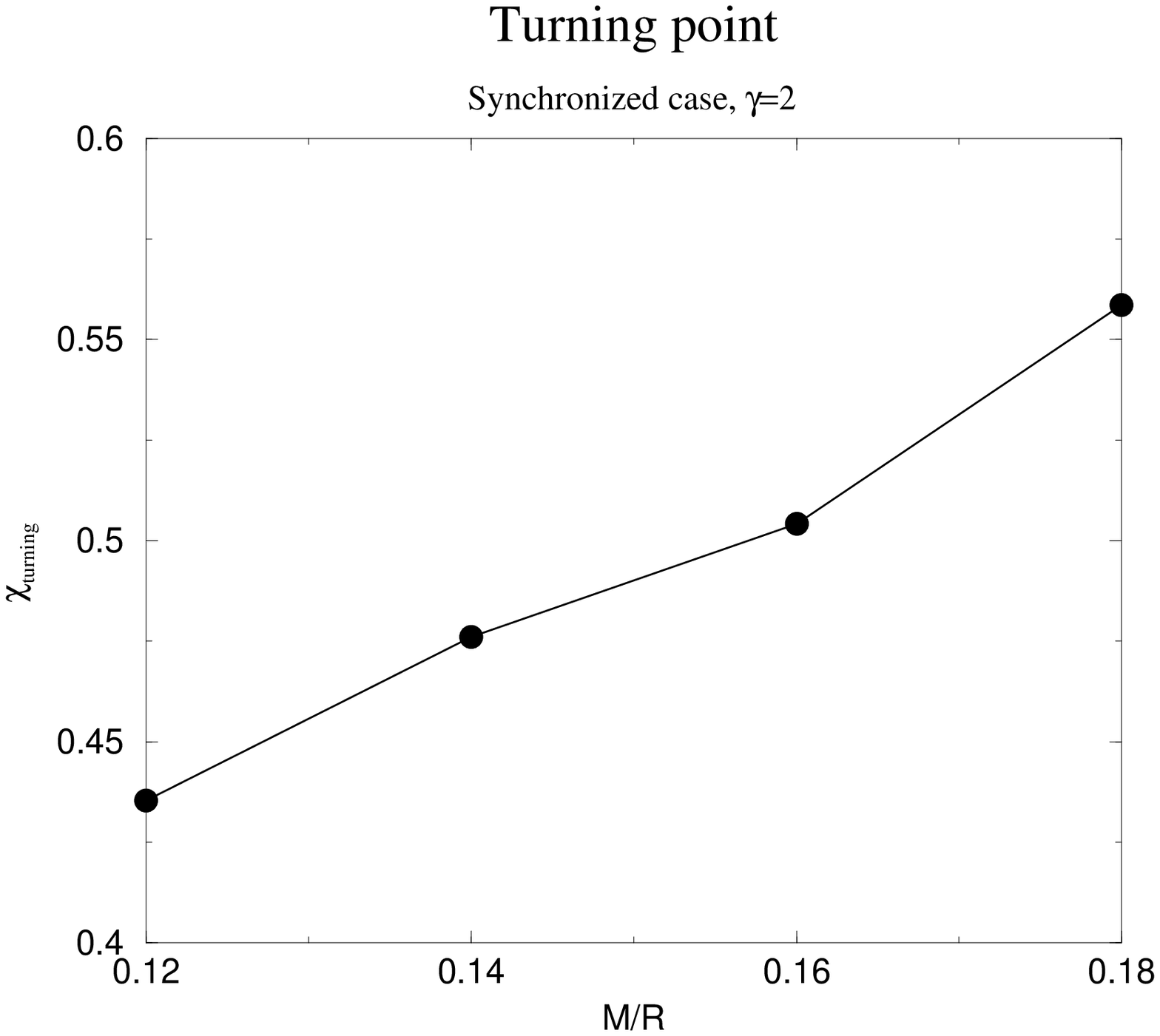}
\end{center}
\caption[]{\label{fig:turning}
Cusp indicator $\chi$ at the turning point of
the ADM mass along a sequence
of synchronized identical mass binaries,
as a function of the compactness $M/R$.
}
\end{figure}%

The relative change in central energy density is presented
as a function of $\chi$ in Fig.~\ref{fig:dec_vs_chi}.
It is clearly seen that the sequences for the synchronized case
with the same compactness coincide with each other,
while those with the different compactness split into curves
with respect to each compactness.
For example, the relative change for the identical binary system
with the compactness $M/R=0.14$ coincides with that
for the more massive star of the different mass binary with
$M/R=0.12$ vs. 0.14 and that for the less massive star with
$M/R=0.14$ vs. 0.16.
However, the sequences for stars with $M/R=0.12$, 0.14, and 0.16
are different from each other.

Finally, let us comment on the behavior of the turning point
of the ADM mass (binding energy) along a sequence.
We never found any turning points for irrotational binary systems
with the polytropic index $\gamma=2$.
On the other hand, it appeared clearly for the synchronized case.
In Fig.~\ref{fig:turning}, the cusp indicator $\chi$ at the turning point
of synchronized identical mass binary systems
is plotted as a function of the compactness $M/R$.
Since the accuracy of the determination of the ADM mass
is order of $10^{-5}$ in our calculations,
the relative errors on the orbital angular velocity
and the quantity $\chi$ at the turning point
become a few percents.
Even when one is taking this fact into account,
it is possible to conclude that
the line in Fig.~\ref{fig:turning} increases
proportionally to the compactness of the stars.
This behavior agrees with the value of $\chi_{\rm turning}$
that we have found in Newtonian calculations ($M/R=0$),
namely $\chi_{\rm turning}=0.2862$ (Table~I of Paper~II).

\section{Summary} \label{s:summary}

In the present paper, we have computed constant baryon number sequences
of binary neutron stars in quasiequilibrium,
in both cases of synchronized and irrotational motion
and in both cases of identical and different mass systems.
We have performed a general relativistic treatment, within
the IWM approximation (conformally flat spatial metric).
We have used a polytropic equation of state with an
adiabatic index $\gamma=2$.

The summary of our results is that
(1) Among the quasiequilibrium sequences we have calculated in the
present article, only the synchronized identical star binaries
terminate by the contact between two stars,
while all the other types of sequences (synchronized different star
binaries, irrotational identical star and different star binaries)
end at the mass shedding points.
Note here that we cannot conclude about the end points of
{\em non-quasiequilibrium} sequences (i.e. dynamical sequences),
such that binary systems without viscosity
but with intrinsic non-aligned spins, those with slight viscosity
and deviating from both the irrotational state and the synchronized one,
those with infalling radial velocity, and so on.
(2) For identical mass binary systems
there is no turning point of the ADM mass
in the irrotational case,
while there is clearly one before the end point of the sequence
in the synchronized case.
(3) It is more difficult to see the turning points of the ADM mass
for different mass binary systems than for identical ones.
(4) The deformation of the star is determined
by the orbital separation and the mass ratio
and is not affected much by its compactness.
(5) The decrease of the central energy density depends
on the compactness of the star and not on that of its companion.

\acknowledgments

KT is grateful to all members of the numerical relativity group
at Albert-Einstein-Institut leaded by E. Seidel
for their hearty hospitality during his stay.
KT also would like to thank K. Uryu for useful discussions.
The code development have been performed
on SGI workstations purchased thanks to a special grant from the C.N.R.S.
The numerical computations were mainly carried out on a SGI Origin2000
at Albert-Einstein-Institut.
KT acknowledges a Grant-in-Aid for Scientific Research (No. 14-06898)
of the Japanese Ministry of Education, Culture, Sports, Science and Technology.


\newpage

\begin{table}
\caption{Orbital angular velocity, ADM mass, total angular momentum,
virial errors, axial ratios, relative change in central energy density,
and the cusp indicator $\chi$ along constant baryon number sequences
of identical mass binaries in synchronized motion, for various
compactness.
The symbol $\dagger$ denotes the turning point of the sequences.
}
 \begin{ruledtabular}
  \begin{tabular}{rccccccccccc}
  \multicolumn{12}{c}{Identical mass stars, Synchronized case} \\
  $\bar{d}_G$&$\tilde{d}$&$\bar{\Omega}$&$\bar{M}$&$\bar{J}$&
  $|\bar{VE(M)}|$&$|\bar{VE(GB)}|$&$|\bar{VE(FUS)}|$&
  $a_2/a_1$&$a_3/a_1$&$\d e_{\rm c}$&$\chi$ \\ \hline
  \multicolumn{12}{c}
  {$M/R=0.12~{\rm vs.}~0.12,~~\bar{M}_{\rm B}=0.1299~{\rm vs.}~0.1299$} \\
  4.873& 2.757& 4.299(-2)& 0.2432& 7.675(-2)&
  8.840(-6)& 1.453(-5)& 1.350(-5)& 0.9846& 0.9773& -1.136(-2)& 0.9395 \\
  3.655& 2.033& 6.473(-2)& 0.2429& 7.057(-2)&
  6.931(-6)& 1.142(-5)& 1.042(-5)& 0.9637& 0.9438& -2.675(-2)& 0.8500 \\
  3.046& 1.639& 8.366(-2)& 0.2427& 6.793(-2)&
  6.181(-6)& 8.837(-6)& 7.788(-6)& 0.9317& 0.8970& -4.736(-2)& 0.7236 \\
  2.864& 1.509& 9.125(-2)& 0.2427& 6.733(-2)&
  4.256(-6)& 7.069(-6)& 6.307(-6)& 0.9126& 0.8713& -5.820(-2)& 0.6531 \\
  2.742& 1.415& 9.701(-2)& 0.2426& 6.703(-2)&
  5.475(-6)& 6.589(-6)& 5.809(-6)& 0.8942& 0.8478& -6.772(-2)& 0.5875 \\
  2.650& 1.338& 1.018(-1)& 0.2426& 6.688(-2)&
  8.194(-7)& 8.531(-6)& 7.661(-6)& 0.8752& 0.8246& -7.666(-2)& 0.5214 \\
$\dagger$  2.565& 1.255& 1.066(-1)& 0.2426& 6.684(-2)&
  8.662(-6)& 2.064(-5)& 1.707(-5)& 0.8492& 0.7949& -8.698(-2)& 0.4353 \\
  2.559& 1.249& 1.070(-1)& 0.2426& 6.684(-2)&
  4.567(-6)& 2.755(-5)& 2.337(-5)& 0.8470& 0.7924& -8.781(-2)& 0.4278 \\
  2.528& 1.215& 1.089(-1)& 0.2426& 6.685(-2)&
  1.195(-6)& 3.725(-5)& 3.236(-5)& 0.8342& 0.7783& -9.223(-2)& 0.3858 \\
  \multicolumn{12}{c}
  {$M/R=0.14~{\rm vs.}~0.14,~~\bar{M}_{\rm B}=0.1461~{\rm vs.}~0.1461$} \\
  4.873& 2.979& 4.505(-2)& 0.2708& 9.090(-2)&
  1.558(-5)& 2.677(-5)& 2.525(-5)& 0.9877& 0.9821& -1.119(-2)& 0.9513 \\
  3.655& 2.211& 6.764(-2)& 0.2704& 8.357(-2)&
  1.271(-5)& 2.189(-5)& 2.026(-5)& 0.9720& 0.9560& -2.588(-2)& 0.8805 \\
  3.046& 1.803& 8.719(-2)& 0.2702& 8.031(-2)&
  9.695(-6)& 1.673(-5)& 1.527(-5)& 0.9494& 0.9209& -4.480(-2)& 0.7846 \\
  2.742& 1.581& 1.008(-1)& 0.2700& 7.902(-2)&
  9.415(-6)& 1.372(-5)& 1.214(-5)& 0.9253& 0.8864& -6.247(-2)& 0.6900 \\
  2.620& 1.485& 1.074(-1)& 0.2700& 7.864(-2)&
  3.142(-6)& 1.728(-5)& 1.568(-5)& 0.9099& 0.8658& -7.263(-2)& 0.6326 \\
  2.499& 1.380& 1.147(-1)& 0.2700& 7.839(-2)&
  1.412(-5)& 2.865(-5)& 2.280(-5)& 0.8878& 0.8378& -8.559(-2)& 0.5538 \\
$\dagger$  2.413& 1.297& 1.203(-1)& 0.2700& 7.833(-2)&
  1.192(-5)& 3.127(-5)& 2.535(-5)& 0.8652& 0.8108& -9.715(-2)& 0.4761 \\
  2.377& 1.258& 1.229(-1)& 0.2700& 7.834(-2)&
  9.280(-6)& 3.548(-5)& 2.930(-5)& 0.8522& 0.7960& -1.030(-1)& 0.4328 \\
  2.346& 1.221& 1.252(-1)& 0.2700& 7.836(-2)&
  1.396(-5)& 7.557(-5)& 6.519(-5)& 0.8390& 0.7814& -1.084(-1)& 0.3893 \\
  \multicolumn{12}{c}
  {$M/R=0.16~{\rm vs.}~0.16,~~\bar{M}_{\rm B}=0.1600~{\rm vs.}~0.1600$} \\
  4.264& 2.829& 5.634(-2)& 0.2935& 9.912(-2)&
  2.509(-5)& 4.471(-5)& 4.242(-5)& 0.9862& 0.9791& -1.699(-2)& 0.9417 \\
  3.655& 2.414& 6.990(-2)& 0.2933& 9.497(-2)&
  2.255(-5)& 4.021(-5)& 3.777(-5)& 0.9786& 0.9662& -2.641(-2)& 0.9059 \\
  3.046& 1.983& 8.992(-2)& 0.2930& 9.114(-2)&
  1.815(-5)& 3.248(-5)& 3.011(-5)& 0.9626& 0.9397& -4.486(-2)& 0.8328 \\
  2.620& 1.659& 1.104(-1)& 0.2928& 8.898(-2)&
  1.382(-5)& 2.393(-5)& 2.181(-5)& 0.9367& 0.9005& -7.066(-2)& 0.7240 \\
  2.438& 1.506& 1.218(-1)& 0.2927& 8.836(-2)&
  2.887(-5)& 4.340(-5)& 3.510(-5)& 0.9153& 0.8708& -8.872(-2)& 0.6414 \\
  2.316& 1.393& 1.306(-1)& 0.2927& 8.813(-2)&
  2.270(-5)& 3.892(-5)& 3.056(-5)& 0.8929& 0.8419& -1.053(-1)& 0.5597 \\
$\dagger$  2.255& 1.331& 1.354(-1)& 0.2927& 8.810(-2)&
  2.050(-5)& 3.994(-5)& 3.153(-5)& 0.8771& 0.8226& -1.156(-1)& 0.5042 \\
  2.194& 1.261& 1.405(-1)& 0.2927& 8.815(-2)&
  3.159(-6)& 6.903(-5)& 5.741(-5)& 0.8556& 0.7975& -1.277(-1)& 0.4312 \\
  2.164& 1.220& 1.432(-1)& 0.2927& 8.821(-2)&
  3.129(-5)& 5.921(-5)& 5.015(-5)& 0.8409& 0.7813& -1.350(-1)& 0.3837 \\
  \multicolumn{12}{c}
  {$M/R=0.18~{\rm vs.}~0.18,~~\bar{M}_{\rm B}=0.1710~{\rm vs.}~0.1710$} \\
  3.655& 2.650& 7.156(-2)& 0.3108& 1.040(-1)&
  3.917(-5) &7.192(-5)& 6.851(-5)& 0.9840& 0.9745& -3.007(-2)& 0.9273 \\
  3.046& 2.188& 9.193(-2)& 0.3104& 9.968(-2)&
  3.292(-5)& 6.062(-5)& 5.706(-5)& 0.9726& 0.9547& -5.015(-2)& 0.8716 \\
  2.559& 1.796& 1.163(-1)& 0.3101& 9.679(-2)&
  2.624(-5)& 4.584(-5)& 4.207(-5)& 0.9512& 0.9201& -8.243(-2)& 0.7743 \\
  2.316& 1.581& 1.330(-1)& 0.3100& 9.582(-2)&
  3.933(-5)& 6.381(-5)& 5.182(-5)& 0.9284& 0.8865& -1.108(-1)& 0.6802 \\
  2.194& 1.463& 1.428(-1)& 0.3100& 9.555(-2)&
  3.324(-5)& 3.124(-5)& 2.506(-5)& 0.9094& 0.8607& -1.308(-1)& 0.6071 \\
$\dagger$  2.134& 1.398& 1.482(-1)& 0.3100& 9.550(-2)&
  2.666(-5)& 6.812(-5)& 5.487(-5)& 0.8962& 0.8436& -1.431(-1)& 0.5585 \\
  2.073& 1.328& 1.540(-1)& 0.3100& 9.554(-2)&
  2.113(-5)& 7.194(-5)& 5.803(-5)& 0.8788& 0.8220& -1.574(-1)& 0.4966 \\
  2.042& 1.290& 1.571(-1)& 0.3100& 9.559(-2)&
  1.005(-7)& 1.085(-4)& 8.977(-5)& 0.8675& 0.8087& -1.656(-1)& 0.4581 \\
  2.012& 1.248& 1.602(-1)& 0.3100& 9.567(-2)&
  4.965(-5)& 1.943(-4)& 1.632(-4)& 0.8537& 0.7930& -1.747(-1)& 0.4120 \\
  \end{tabular}
 \end{ruledtabular}
 \label{table1}
\end{table}%

\begin{table}
\caption{Same as Table~\ref{table1} but for different mass binary
systems. The superscript `` ' '' denotes the values for
the more massive star.
}
 \begin{ruledtabular}
  \begin{tabular}{rccccccccccc}
  \multicolumn{12}{c}{Different mass stars, Synchronized case} \\
  $\bar{d}_G$&$\tilde{d}$&$\bar{\Omega}$&$\bar{M}$&$\bar{J}$&
  $|\bar{VE(M)}|$&$|\bar{VE(GB)}|$&$|\bar{VE(FUS)}|$&
  $a_2/a_1$&$a_3/a_1$&$\d e_{\rm c}$&$\chi$ \\
   & & & & & & & &$a_2'/a_1'$&$a_3'/a_1'$&$\d e_{\rm c}'$&$\chi'$  \\ \hline
  \multicolumn{12}{c}
  {$M/R=0.12~{\rm vs.}~0.14,~~\bar{M}_{\rm B}=0.1299~{\rm vs.}~0.1461$} \\
  4.873& 2.863& 4.404(-2)& 0.2570& 8.347(-2)&
  1.201(-5)& 2.022(-5)& 1.895(-5)& 0.9827& 0.9754& -1.197(-2)& 0.9345 \\
   & & & & & & &                 & 0.9891& 0.9833& -1.066(-2)& 0.9548 \\
  3.655& 2.117& 6.621(-2)& 0.2567& 7.675(-2)&
  9.690(-6)& 1.633(-5)& 1.501(-5)& 0.9593& 0.9391& -2.824(-2)& 0.8376 \\
   & & & & & & &                 & 0.9750& 0.9592& -2.467(-2)& 0.8891 \\
  3.046& 1.715& 8.547(-2)& 0.2565& 7.382(-2)&
  7.292(-6)& 1.220(-5)& 1.106(-5)& 0.9234& 0.8882& -5.023(-2)& 0.7001 \\
   & & & & & & &                 & 0.9547& 0.9265& -4.264(-2)& 0.8001 \\
  2.864& 1.583& 9.317(-2)& 0.2564& 7.312(-2)&
  6.226(-6)& 1.043(-5)& 9.430(-6)& 0.9018& 0.8600& -6.189(-2)& 0.6227 \\
   & & & & & & &                 & 0.9434& 0.9095& -5.169(-2)& 0.7532 \\
  2.742& 1.488& 9.899(-2)& 0.2564& 7.275(-2)&
  1.408(-5)& 1.812(-5)& 1.523(-5)& 0.8807& 0.8338& -7.209(-2)& 0.5501 \\
   & & & & & & &                 & 0.9330& 0.8946& -5.930(-2)& 0.7121 \\
  2.681& 1.437& 1.022(-1)& 0.2563& 7.261(-2)&
  1.605(-5)& 2.080(-5)& 1.631(-5)& 0.8665& 0.8170& -7.840(-2)& 0.5026 \\
   & & & & & & &                 & 0.9265& 0.8856& -6.392(-2)& 0.6870 \\
  2.620& 1.383& 1.055(-1)& 0.2563& 7.249(-2)&
  1.185(-5)& 2.514(-5)& 2.050(-5)& 0.8486& 0.7965& -8.565(-2)& 0.4439 \\
   & & & & & & &                 & 0.9190& 0.8753& -6.900(-2)& 0.6582 \\
  2.577& 1.342& 1.080(-1)& 0.2563& 7.243(-2)&
  1.161(-6)& 4.051(-5)& 3.446(-5)& 0.8327& 0.7789& -9.144(-2)& 0.3923 \\
   & & & & & & &                 & 0.9128& 0.8671& -7.294(-2)& 0.6352 \\
  \multicolumn{12}{c}
  {$M/R=0.14~{\rm vs.}~0.16,~~\bar{M}_{\rm B}=0.1461~{\rm vs.}~0.1600$} \\
  4.873& 3.102& 4.586(-2)& 0.2823& 9.688(-2)&
  2.076(-5)& 3.649(-5)& 3.473(-5)& 0.9865& 0.9810& -1.164(-2)& 0.9484 \\
   & & & & & & &                 & 0.9912& 0.9869& -1.119(-2)& 0.9635 \\
  3.655& 2.307& 6.879(-2)& 0.2818& 8.905(-2)&
  1.736(-5)& 3.055(-5)& 2.855(-5)& 0.9694& 0.9534& -2.691(-2)& 0.8733 \\
   & & & & & & &                 & 0.9805& 0.9679& -2.550(-2)& 0.9110 \\
  3.046& 1.888& 8.858(-2)& 0.2816& 8.552(-2)&
  1.370(-5)& 2.420(-5)& 2.232(-5)& 0.9449& 0.9161& -4.664(-2)& 0.7717 \\
   & & & & & & &                 & 0.9657& 0.9429& -4.331(-2)& 0.8417 \\
  2.742& 1.662& 1.024(-1)& 0.2814& 8.408(-2)&
  1.132(-5)& 1.941(-5)& 1.771(-5)& 0.9186& 0.8794& -6.518(-2)& 0.6711 \\
   & & & & & & &                 & 0.9507& 0.9193& -5.931(-2)& 0.7762 \\
  2.620& 1.564& 1.090(-1)& 0.2814& 8.364(-2)&
  2.342(-5)& 3.453(-5)& 2.770(-5)& 0.9016& 0.8572& -7.584(-2)& 0.6096 \\
   & & & & & & &                 & 0.9416& 0.9056& -6.817(-2)& 0.7384 \\
  2.499& 1.460& 1.163(-1)& 0.2814& 8.329(-2)&
  2.577(-5)& 3.575(-5)& 2.861(-5)& 0.8774& 0.8271& -8.945(-2)& 0.5251 \\
   & & & & & & &                 & 0.9295& 0.8882& -7.911(-2)& 0.6897 \\
  2.438& 1.403& 1.203(-1)& 0.2813& 8.318(-2)&
  1.268(-5)& 4.801(-5)& 3.990(-5)& 0.8603& 0.8071& -9.800(-2)& 0.4677 \\
   & & & & & & &                 & 0.9217& 0.8775& -8.567(-2)& 0.6595 \\
  2.377& 1.341& 1.245(-1)& 0.2813& 8.310(-2)&
  1.713(-5)& 1.014(-4)& 8.814(-5)& 0.8376& 0.7814& -1.079(-1)& 0.3927 \\
   & & & & & & &                 & 0.9124& 0.8650& -9.308(-2)& 0.6238 \\
  \multicolumn{12}{c}
  {$M/R=0.16~{\rm vs.}~0.18,~~\bar{M}_{\rm B}=0.1600~{\rm vs.}~0.1710$} \\
  4.264& 2.957& 5.704(-2)& 0.3023& 1.037(-1)&
  3.359(-5)& 6.089(-5)& 5.824(-5)& 0.9852& 0.9782& -1.745(-2)& 0.9393 \\
   & & & & & & &                 & 0.9901& 0.9848& -1.904(-2)& 0.9566 \\
  3.655& 2.526& 7.074(-2)& 0.3020& 9.936(-2)&
  3.049(-5)& 5.543(-5)& 5.256(-5)& 0.9772& 0.9648& -2.712(-2)& 0.9021 \\
   & & & & & & &                 & 0.9850& 0.9754& -2.934(-2)& 0.9301 \\
  3.046& 2.080& 9.094(-2)& 0.3017& 9.529(-2)&
  2.522(-5)& 4.605(-5)& 4.311(-5)& 0.9602& 0.9373& -4.608(-2)& 0.8260 \\
   & & & & & & &                 & 0.9743& 0.9564& -4.896(-2)& 0.8764 \\
  2.620& 1.747& 1.116(-1)& 0.3015& 9.293(-2)&
  1.940(-5)& 3.529(-5)& 3.267(-5)& 0.9328& 0.8963& -7.264(-2)& 0.7127 \\
   & & & & & & &                 & 0.9577& 0.9289& -7.522(-2)& 0.7990 \\
  2.438& 1.592& 1.230(-1)& 0.3014& 9.221(-2)&
  3.520(-5)& 5.539(-5)& 4.539(-5)& 0.9100& 0.8653& -9.132(-2)& 0.6266 \\
   & & & & & & &                 & 0.9448& 0.9089& -9.288(-2)& 0.7431 \\
  2.316& 1.479& 1.318(-1)& 0.3013& 9.188(-2)&
  3.541(-5)& 5.009(-5)& 4.242(-5)& 0.8862& 0.8350& -1.085(-1)& 0.5411 \\
   & & & & & & &                 & 0.9324& 0.8907& -1.081(-1)& 0.6915 \\
  2.237& 1.398& 1.381(-1)& 0.3013& 9.177(-2)&
  1.418(-5)& 7.574(-5)& 6.303(-5)& 0.8633& 0.8075& -1.226(-1)& 0.4625 \\
   & & & & & & &                 & 0.9214& 0.8753& -1.204(-1)& 0.6480 \\
  2.194& 1.350& 1.417(-1)& 0.3013& 9.174(-2)&
  1.249(-5)& 1.266(-4)& 1.084(-4)& 0.8465& 0.7883& -1.315(-1)& 0.4063 \\
   & & & & & & &                 & 0.9141& 0.8655& -1.278(-1)& 0.6200 \\
  \end{tabular}
 \end{ruledtabular}
 \label{table2}
\end{table}%

\begin{table}
\caption{Same as Table~\ref{table1}, but for irrotational binaries.
}
 \begin{ruledtabular}
  \begin{tabular}{rccccccccccc}
  \multicolumn{12}{c}{Identical mass stars, Irrotational case} \\
  $\bar{d}_G$&$\tilde{d}$&$\bar{\Omega}$&$\bar{M}$&$\bar{J}$&
  $|\bar{VE(M)}|$&$|\bar{VE(GB)}|$&$|\bar{VE(FUS)}|$&
  $a_2/a_1$&$a_3/a_1$&$\d e_{\rm c}$&$\chi$ \\ \hline
  \multicolumn{12}{c}
  {$M/R=0.12~{\rm vs.}~0.12,~~\bar{M}_{\rm B}=0.1299~{\rm vs.}~0.1299$} \\
  6.090& 3.472& 3.120(-2)& 0.2434& 8.076(-2)&
  9.264(-6)& 1.530(-5)& 1.439(-5)& 0.9916& 0.9945& -8.061(-5)& 0.9837 \\
  4.872& 2.780& 4.304(-2)& 0.2432& 7.365(-2)&
  8.394(-6)& 1.371(-5)& 1.266(-5)& 0.9846& 0.9882& -2.038(-4)& 0.9654 \\
  3.653& 2.072& 6.488(-2)& 0.2428& 6.589(-2)&
  5.992(-6)& 9.535(-6)& 8.445(-6)& 0.9631& 0.9681& -8.804(-4)& 0.9055 \\
  3.044& 1.694& 8.397(-2)& 0.2425& 6.180(-2)&
  3.377(-6)& 4.933(-6)& 3.962(-6)& 0.9299& 0.9369& -2.613(-3)& 0.8082 \\
  2.861& 1.570& 9.164(-2)& 0.2423& 6.058(-2)&
  6.164(-6)& 8.369(-6)& 6.394(-6)& 0.9100& 0.9183& -3.929(-3)& 0.7459 \\
  2.738& 1.478& 9.746(-2)& 0.2423& 5.978(-2)&
  4.595(-6)& 6.938(-6)& 5.079(-6)& 0.8896& 0.8988& -5.328(-3)& 0.6692 \\
  2.616& 1.375& 1.040(-1)& 0.2422& 5.900(-2)&
  1.337(-5)& 1.239(-5)& 7.816(-6)& 0.8582& 0.8705& -7.548(-3)& 0.5655 \\
  2.554& 1.304& 1.076(-1)& 0.2422& 5.864(-2)&
  5.612(-7)& 1.134(-6)& 6.392(-6)& 0.8283& 0.8428& -9.566(-3)& 0.4476 \\
  2.523& 1.259& 1.095(-1)& 0.2421& 5.846(-2)&
  2.005(-5)& 5.038(-6)& 1.880(-5)& 0.8064& 0.8223& -1.111(-2)& 0.3720 \\
  \multicolumn{12}{c}
  {$M/R=0.14~{\rm vs.}~0.14,~~\bar{M}_{\rm B}=0.1461~{\rm vs.}~0.1461$} \\
  4.872& 3.001& 4.510(-2)& 0.2707& 8.755(-2)&
  1.494(-5)& 2.565(-5)& 2.409(-5)& 0.9877& 0.9914& -1.870(-4)& 0.9736 \\
  3.653& 2.249& 6.780(-2)& 0.2702& 7.858(-2)&
  1.136(-5)& 1.924(-5)& 1.748(-5)& 0.9717& 0.9766& -7.114(-4)& 0.9284 \\
  3.044& 1.856& 8.753(-2)& 0.2699& 7.384(-2)&
  7.459(-6)& 1.222(-5)& 1.054(-5)& 0.9483& 0.9545& -1.940(-3)& 0.8594 \\
  2.739& 1.645& 1.014(-1)& 0.2697& 7.146(-2)&
  5.141(-6)& 6.971(-6)& 5.511(-6)& 0.9231& 0.9309& -3.673(-3)& 0.7812 \\
  2.617& 1.553& 1.080(-1)& 0.2696& 7.053(-2)&
  8.504(-6)& 1.243(-5)& 9.189(-6)& 0.9067& 0.9154& -5.005(-3)& 0.7255 \\
  2.495& 1.453& 1.154(-1)& 0.2695& 6.962(-2)&
  1.532(-5)& 2.005(-5)& 1.304(-5)& 0.8819& 0.8929& -6.999(-3)& 0.6432 \\
  2.433& 1.392& 1.195(-1)& 0.2694& 6.918(-2)&
  1.200(-5)& 1.575(-5)& 7.702(-6)& 0.8616& 0.8742& -8.502(-3)& 0.5623 \\
  2.403& 1.353& 1.216(-1)& 0.2694& 6.896(-2)&
  5.376(-6)& 1.547(-6)& 1.295(-5)& 0.8453& 0.8591& -9.722(-3)& 0.4950 \\
  2.372& 1.305& 1.238(-1)& 0.2694& 6.874(-2)&
  6.626(-5)& 6.375(-5)& 8.573(-5)& 0.8227& 0.8382& -1.187(-2)& 0.4140 \\
  \multicolumn{12}{c}
  {$M/R=0.16~{\rm vs.}~0.16,~~\bar{M}_{\rm B}=0.1600~{\rm vs.}~0.1600$} \\
  4.263& 2.858& 5.643(-2)& 0.2934& 9.499(-2)&
  2.404(-5)& 4.283(-5)& 4.043(-5)& 0.9863& 0.9903& -3.097(-4)& 0.9689 \\
  3.653& 2.452& 7.007(-2)& 0.2931& 8.988(-2)&
  2.093(-5)& 3.718(-5)& 3.456(-5)& 0.9786& 0.9833& -6.159(-4)& 0.9468 \\
  3.044& 2.037& 9.028(-2)& 0.2927& 8.461(-2)&
  1.544(-5)& 2.709(-5)& 2.441(-5)& 0.9621& 0.9679& -1.548(-3)& 0.8975 \\
  2.739& 1.821& 1.044(-1)& 0.2924& 8.194(-2)&
  1.117(-5)& 1.921(-5)& 1.659(-5)& 0.9454& 0.9522& -2.795(-3)& 0.8464 \\
  2.556& 1.684& 1.147(-1)& 0.2922& 8.036(-2)&
  1.693(-5)& 1.689(-5)& 1.253(-5)& 0.9291& 0.9368& -4.126(-3)& 0.7937 \\
  2.434& 1.585& 1.227(-1)& 0.2921& 7.933(-2)&
  1.539(-5)& 2.183(-5)& 1.667(-5)& 0.9127& 0.9213& -5.743(-3)& 0.7351 \\
  2.312& 1.478& 1.316(-1)& 0.2920& 7.833(-2)&
  1.866(-5)& 2.815(-5)& 1.727(-5)& 0.8876& 0.8986& -8.268(-3)& 0.6512 \\
  2.251& 1.405& 1.365(-1)& 0.2919& 7.783(-2)&
  1.504(-5)& 1.274(-5)& 2.947(-5)& 0.8614& 0.8745& -1.052(-2)& 0.5389 \\
  2.220& 1.355& 1.390(-1)& 0.2919& 7.754(-2)&
  1.168(-4)& 1.423(-4)& 1.751(-4)& 0.8386& 0.8535& -1.314(-2)& 0.4546 \\
  \multicolumn{12}{c}
  {$M/R=0.18~{\rm vs.}~0.18,~~\bar{M}_{\rm B}=0.1710~{\rm vs.}~0.1710$} \\
  3.654& 2.692& 7.173(-2)& 0.3106& 9.902(-2)&
  3.784(-5)& 6.963(-5)& 6.598(-5)& 0.9842& 0.9887& -6.028(-4)& 0.9618 \\
  3.044& 2.247& 9.228(-2)& 0.3101& 9.334(-2)&
  3.055(-5)& 5.616(-5)& 5.218(-5)& 0.9728& 0.9781& -1.402(-3)& 0.9269 \\
  2.740& 2.018& 1.066(-1)& 0.3099& 9.046(-2)&
  2.478(-5)& 4.546(-5)& 4.136(-5)& 0.9617& 0.9677& -2.408(-3)& 0.8923 \\
  2.557& 1.877& 1.170(-1)& 0.3097& 8.874(-2)&
  2.091(-5)& 3.724(-5)& 3.304(-5)& 0.9513& 0.9579& -3.502(-3)& 0.8597 \\
  2.313& 1.678& 1.340(-1)& 0.3094& 8.650(-2)&
  2.473(-5)& 4.064(-5)& 3.310(-5)& 0.9287& 0.9367& -6.232(-3)& 0.7855 \\
  2.191& 1.570& 1.441(-1)& 0.3092& 8.542(-2)&
  3.136(-5)& 5.104(-5)& 3.558(-5)& 0.9091& 0.9188& -8.975(-3)& 0.7201 \\
  2.130& 1.508& 1.497(-1)& 0.3092& 8.490(-2)&
  2.370(-5)& 3.947(-5)& 2.302(-5)& 0.8934& 0.9044& -1.088(-2)& 0.6571 \\
  2.099& 1.466& 1.525(-1)& 0.3091& 8.461(-2)&
  1.552(-5)& 1.699(-5)& 3.932(-5)& 0.8783& 0.8905& -1.254(-2)& 0.5868 \\
  2.068& 1.411& 1.553(-1)& 0.3091& 8.421(-2)&
  1.832(-4)& 2.664(-4)& 3.122(-4)& 0.8547& 0.8691& -1.612(-2)& 0.4960 \\
  \end{tabular}
 \end{ruledtabular}
 \label{table3}
\end{table}%

\begin{table}
\caption{Same as Table~\ref{table3} but for different mass binary
systems. The superscript `` ' '' denotes the values for
the more massive star.
}
 \begin{ruledtabular}
  \begin{tabular}{rccccccccccc}
  \multicolumn{12}{c}{Different mass stars, Irrotational case} \\
  $\bar{d}_G$&$\tilde{d}$&$\bar{\Omega}$&$\bar{M}$&$\bar{J}$&
  $|\bar{VE(M)}|$&$|\bar{VE(GB)}|$&$|\bar{VE(FUS)}|$&
  $a_2/a_1$&$a_3/a_1$&$\d e_{\rm c}$&$\chi$ \\
   & & & & & & & &$a_2'/a_1'$&$a_3'/a_1'$&$\d e_{\rm c}'$&$\chi'$  \\ \hline
  \multicolumn{12}{c}
  {$M/R=0.12~{\rm vs.}~0.14,~~\bar{M}_{\rm B}=0.1299~{\rm vs.}~0.1299$} \\
  6.090& 3.602& 3.199(-2)& 0.2572& 8.790(-2)&
  1.249(-5)& 2.119(-5)& 2.011(-5)& 0.9904& 0.9938& -1.076(-4)& 0.9819 \\
   & & & & & & &                 & 0.9940& 0.9964& -5.955(-5)& 0.9889 \\
  3.653& 2.156& 6.637(-2)& 0.2565& 7.191(-2)&
  8.519(-6)& 1.408(-5)& 1.267(-5)& 0.9586& 0.9645& -1.116(-3)& 0.8950 \\
   & & & & & & &                 & 0.9748& 0.9789& -5.560(-4)& 0.9355 \\
  3.044& 1.769& 8.579(-2)& 0.2562& 6.751(-2)&
  5.310(-6)& 8.355(-6)& 7.043(-6)& 0.9212& 0.9295& -3.284(-3)& 0.7858 \\
   & & & & & & &                 & 0.9538& 0.9591& -1.541(-3)& 0.8732 \\
  2.861& 1.643& 9.359(-2)& 0.2561& 6.619(-2)&
  8.448(-6)& 1.253(-5)& 9.948(-6)& 0.8985& 0.9081& -4.941(-3)& 0.7131 \\
   & & & & & & &                 & 0.9420& 0.9480& -2.256(-3)& 0.8374 \\
  2.739& 1.553& 9.949(-2)& 0.2560& 6.533(-2)&
  1.412(-5)& 1.903(-5)& 1.371(-5)& 0.8750& 0.8867& -6.731(-3)& 0.6375 \\
   & & & & & & &                 & 0.9311& 0.9379& -2.988(-3)& 0.8038 \\
  2.677& 1.503& 1.027(-1)& 0.2559& 6.490(-2)&
  1.371(-5)& 1.827(-5)& 1.285(-5)& 0.8583& 0.8712& -7.965(-3)& 0.5768 \\
   & & & & & & &                 & 0.9243& 0.9316& -3.465(-3)& 0.7824 \\
  2.616& 1.446& 1.061(-1)& 0.2559& 6.448(-2)&
  9.790(-6)& 1.443(-5)& 8.245(-6)& 0.8333& 0.8481& -9.742(-3)& 0.4796 \\
   & & & & & & &                 & 0.9163& 0.9242& -4.057(-3)& 0.7563 \\
  2.585& 1.412& 1.079(-1)& 0.2559& 6.428(-2)&
  6.266(-7)& 5.076(-6)& 3.032(-6)& 0.8146& 0.8308& -1.118(-2)& 0.4100 \\
   & & & & & & &                 & 0.9115& 0.9197& -4.414(-3)& 0.7404 \\
  \multicolumn{12}{c}
  {$M/R=0.14~{\rm vs.}~0.16,~~\bar{M}_{\rm B}=0.1461~{\rm vs.}~0.1600$} \\
  4.872& 3.125& 4.591(-2)& 0.2822& 9.348(-2)&
  2.010(-5)& 3.534(-5)& 3.351(-5)& 0.9865& 0.9907& -2.313(-4)& 0.9714 \\
   & & & & & & &                 & 0.9913& 0.9944& -1.466(-4)& 0.9819 \\
  3.653& 2.346& 6.896(-2)& 0.2817& 8.401(-2)&
  1.587(-5)& 2.772(-5)& 2.558(-5)& 0.9691& 0.9746& -8.582(-4)& 0.9225 \\
   & & & & & & &                 & 0.9805& 0.9846& -5.065(-4)& 0.9509 \\
  3.044& 1.941& 8.893(-2)& 0.2813& 7.901(-2)&
  1.123(-5)& 1.927(-5)& 1.713(-5)& 0.9436& 0.9507& -2.314(-3)& 0.8477 \\
   & & & & & & &                 & 0.9653& 0.9703& -1.292(-3)& 0.9052 \\
  2.739& 1.726& 1.029(-1)& 0.2810& 7.650(-2)&
  1.417(-5)& 2.285(-5)& 1.876(-5)& 0.9161& 0.9249& -4.405(-3)& 0.7630 \\
   & & & & & & &                 & 0.9498& 0.9558& -2.366(-3)& 0.8577 \\
  2.617& 1.634& 1.096(-1)& 0.2809& 7.551(-2)&
  1.342(-5)& 2.021(-5)& 1.619(-5)& 0.8974& 0.9072& -5.894(-3)& 0.6942 \\
   & & & & & & &                 & 0.9405& 0.9470& -3.135(-3)& 0.8285 \\
  2.495& 1.534& 1.171(-1)& 0.2808& 7.454(-2)&
  1.867(-5)& 2.751(-5)& 1.896(-5)& 0.8692& 0.8817& -8.357(-3)& 0.6020 \\
   & & & & & & &                 & 0.9278& 0.9353& -4.218(-3)& 0.7882 \\
  2.434& 1.472& 1.211(-1)& 0.2808& 7.406(-2)&
  6.809(-6)& 1.456(-5)& 4.008(-6)& 0.8423& 0.8569& -1.046(-2)& 0.4925 \\
   & & & & & & &                 & 0.9195& 0.9276& -4.967(-3)& 0.7611 \\
  2.403& 1.434& 1.232(-1)& 0.2808& 7.382(-2)&
  2.180(-5)& 1.778(-5)& 3.298(-5)& 0.8210& 0.8372& -1.249(-2)& 0.4165 \\
   & & & & & & &                 & 0.9146& 0.9230& -5.427(-3)& 0.7442 \\
  \multicolumn{12}{c}
  {$M/R=0.16~{\rm vs.}~0.18,~~\bar{M}_{\rm B}=0.1600~{\rm vs.}~0.1710$} \\
  4.263& 2.988& 5.713(-2)& 0.3022& 9.963(-2)&
  3.261(-5)& 5.931(-5)& 5.653(-5)& 0.9853& 0.9897& -3.607(-4)& 0.9671 \\
   & & & & & & &                 & 0.9904& 0.9939& -2.793(-4)& 0.9789 \\
  3.044& 2.136& 9.130(-2)& 0.3014& 8.885(-2)&
  2.260(-5)& 4.105(-5)& 3.774(-5)& 0.9597& 0.9660& -1.765(-3)& 0.8917 \\
   & & & & & & &                 & 0.9745& 0.9793& -1.225(-3)& 0.9308 \\
  2.618& 1.821& 1.123(-1)& 0.3010& 8.498(-2)&
  1.484(-5)& 2.691(-5)& 2.391(-5)& 0.9312& 0.9393& -4.138(-3)& 0.8036 \\
   & & & & & & &                 & 0.9578& 0.9636& -2.732(-3)& 0.8786 \\
  2.435& 1.676& 1.239(-1)& 0.3008& 8.336(-2)&
  3.135(-5)& 5.044(-5)& 3.855(-5)& 0.9072& 0.9171& -6.569(-3)& 0.7268 \\
   & & & & & & &                 & 0.9448& 0.9516& -4.189(-3)& 0.8376 \\
  2.312& 1.568& 1.328(-1)& 0.3007& 8.230(-2)&
  2.650(-5)& 4.263(-5)& 2.981(-5)& 0.8791& 0.8912& -9.320(-3)& 0.6225 \\
   & & & & & & &                 & 0.9323& 0.9400& -5.731(-3)& 0.7972 \\
  2.282& 1.536& 1.352(-1)& 0.3007& 8.204(-2)&
  1.853(-5)& 3.204(-5)& 1.787(-5)& 0.8661& 0.8793& -1.047(-2)& 0.5663 \\
   & & & & & & &                 & 0.9284& 0.9363& -6.185(-3)& 0.7842 \\
  2.251& 1.498& 1.377(-1)& 0.3006& 8.177(-2)&
  1.672(-5)& 1.483(-5)& 3.366(-5)& 0.8461& 0.8609& -1.245(-2)& 0.4852 \\
   & & & & & & &                 & 0.9240& 0.9322& -6.763(-3)& 0.7692 \\
  2.233& 1.473& 1.392(-1)& 0.3006& 8.159(-2)&
  6.054(-5)& 7.237(-5)& 9.823(-5)& 0.8318& 0.8478& -1.447(-2)& 0.4392 \\
   & & & & & & &                 & 0.9210& 0.9294& -7.143(-3)& 0.7589 \\
  \end{tabular}
 \end{ruledtabular}
 \label{table4}
\end{table}%

\end{document}